\documentclass[aps,pra,reprint,twocolumn,showpacs,floatfix,superscriptaddress]{revtex4-1}

\usepackage{amssymb,amsmath,amstext}                %%   American Physical Society math etc extensions
\usepackage{graphicx}       \graphicspath{ {arxiv/}}                                          %%   Include figure files                                                                                     %%   Include figure files                                            
\usepackage{epstopdf}                                               %%   help with eps -> pdf 
\usepackage{color}                                                     %%   change color
\usepackage{bm}                                                        %%   bold math                                    
\usepackage{appendix}                                              %%   formating appendicies
\usepackage[utf8]{inputenc}
\usepackage{latexsym}
\usepackage{xcolor}
\usepackage{braket}
\usepackage{ulem}
\usepackage{siunitx}
\usepackage{umoline}
\usepackage{epsfig}
\usepackage{graphics}
\usepackage{psfrag}
\usepackage{mathtools}
\usepackage[english]{babel}
\usepackage[T1]{fontenc}
\usepackage{mathtools}

\usepackage{lmodern}
\usepackage{rotating}
\usepackage{multirow}

\usepackage{comment}

\definecolor{lblue} {RGB}{51,71,158}
\usepackage[colorlinks=true,citecolor=blue,linkcolor=blue,urlcolor=lblue]{hyperref}

\usepackage{appendix}

\begin{document}
\setcounter{section}{0}
\title{False signatures of non-ergodic behavior in disordered quantum many-body systems}
\author{Adith Sai Aramthottil}
\affiliation{Szko\l{}a Doktorska Nauk \'Scis\l{}ych i Przyrodniczych, Uniwersytet Jagiello\'nski,  \L{}ojasiewicza 11, PL-30-348 Krak\'ow, Poland}
\affiliation{Instytut Fizyki Teoretycznej, Wydzia\l{} Fizyki, Astronomii i Informatyki Stosowanej,
Uniwersytet Jagiello\'nski,  \L{}ojasiewicza 11, PL-30-348 Krak\'ow, Poland}

\author{Ali Emami Kopaei}
\affiliation{Szko\l{}a Doktorska Nauk \'Scis\l{}ych i Przyrodniczych, Uniwersytet Jagiello\'nski,  \L{}ojasiewicza 11, PL-30-348 Krak\'ow, Poland}
\affiliation{Instytut Fizyki Teoretycznej, Wydzia\l{} Fizyki, Astronomii i Informatyki Stosowanej,
Uniwersytet Jagiello\'nski,  \L{}ojasiewicza 11, PL-30-348 Krak\'ow, Poland}

\author{Piotr Sierant}
\affiliation{Barcelona Supercomputing Center, Barcelona 08034, Spain}

\author{Lev Vidmar}
\affiliation{Department of Theoretical Physics, J. Stefan Institute, SI-1000 Ljubljana, Slovenia}
\affiliation{Department of Physics, Faculty of Mathematics and Physics, University of Ljubljana, SI-1000 Ljubljana, Slovenia}

\author{Jakub Zakrzewski} 
\email{jakub.zakrzewski@uj.edu.pl}
\affiliation{Instytut Fizyki Teoretycznej, Wydzia\l{} Fizyki, Astronomii i Informatyki Stosowanej,
Uniwersytet Jagiello\'nski,  \L{}ojasiewicza 11, PL-30-348 Krak\'ow, Poland}
\affiliation{Mark Kac Complex Systems Research Center, Uniwersytet Jagiello{\'n}ski, Krak{\'o}w, Poland}

\date{\today}

\begin{abstract}
Ergodic isolated quantum many-body systems satisfy the eigenstate thermalization hypothesis (ETH), i.e., the expectation values of local observables in the system's eigenstates approach the predictions of the microcanonical ensemble. However, the ETH does not specify what happens to expectation values of local observables within an energy window 
when the average over disorder realizations is taken. As a result, the expectation values of local observables can be distributed over a relatively wide interval and may exhibit nontrivial structure, as shown in 
[Phys. Rev. B \textbf{104}, 214201 (2021)]
for a quasiperiodic disordered system for site-resolved magnetization. We argue that the non-Gaussian form of this distribution may \textit{falsely} suggest non-ergodicity and a breakdown of ETH. By considering various types of disorder, we find that the functional forms of the distributions of matrix elements of the site-resolved magnetization operator mirror the distribution of the onsite disorder. We argue that this distribution is a direct consequence of the local observable having a finite overlap with moments of the Hamiltonian. We then demonstrate how to adjust the energy window when analyzing expectation values of local observables in disordered quantum many-body systems to correctly assess the system's adherence to ETH, and provide a link between the distribution of expectation values in eigenstates and the outcomes of quench experiments.
\end{abstract}

\maketitle
\section{Introduction}
The fundamental postulates of thermodynamics for closed systems are a consequence of ergodicity, wherein a system explores all accessible microstates within its phase space over long times, allowing physical observables to be determined as ensemble averages with respect to appropriate statistical ensembles~\cite{Huang08,Pathria16}. 
While the attempts to define ergodicity of quantum systems date back to the early days of quantum mechanics and von Neumann's ergodic theorem~\cite{vonNeumann10}, the definition of ergodicity of quantum many-body systems, in the form of the eigenstate thermalization hypothesis (ETH), was developed only several decades later~\cite{Dalessio16}.
Initially, the progress was restricted to semi-classical systems exhibiting classical chaos. In such systems, the eigenvalue statistics were argued to follow the predictions of random matrix theory \cite{Berry77, Casati80, Bohigas84}, in contrast to the Poissonian statistics observed in classically integrable systems.

A leap in our understanding of quantum ergodicity for closed systems was made in \cite{Deutsch91}, where considering a simple Hamiltonian with many disconnected sectors perturbed by a random matrix coupling led to mixing of a large number of states within a small energy range. This resulted in ergodic eigenstates that have expectations similar to those of the microcanonical ensemble.
The questions of quantum ergodicity were considered also by \cite{Peres84, Feingold84,Jensen85, Feingold86}, which led to the introduction~\cite{Srednicki94,Srednicki99,Rigol08} of the eigenstate thermalization hypothesis (ETH) ansatz 
\begin{equation}
\label{eq:ETH}
    \bra{n}\hat{\mathcal{O}}\ket{m}= \mathcal{O}(\overline{E})\delta_{mn}+e^{-S(\overline{E})/2}f(\overline{E},\omega)R_{mn},
\end{equation}
where $\ket{m},\ket{n}$ are the eigenstates of the given Hamiltonian and $E_n$ and $E_m$ are the corresponding eigenvalues, $\overline{E}=(E_m+E_n)/2, \omega =E_m-E_n$. Here, $\mathcal{O}(\overline{E})$ is the expectation value of the microcanonical ensemble, while $f(\overline{E},\omega)$ controls the functional form of thermalization; smooth functions approximate both. For time-reversal symmetry, $R_{mn}$ is a random real variable with zero mean and unit variance $\overline{R_{mn}^2}=1$, while $S(\bar E)$ is the thermodynamic entropy at energy $\bar E$, i.e., the logarithm of the density of states \cite{Burke23}.
Numerical investigations have extensively verified the ETH, ranging from the smoothness of the expectation value of the operator to the exponential decay in off-diagonal elements of the operator with system size, across a diverse range of nonintegrable quantum many-body systems~\cite{Rigol08,Rigol09, Santos10,Ikeda13Finite,Steinigeweg13, Khatami13Fluctuation, Beugeling14,Sorg14, Steinigeweg14, Mondaini15, Beugeling15OffDiagonal,Mondaini16,Yoshizawa18, Jansen19, Schonle21, Noh23,
LeBlond19,  LeBlond20,  Richter20, Brenes20, Brenes20a}. Recently, the notions of quantum ergodicity and ETH were connected~\cite{Pappalardi22,Pappalardi23,pappalardi2023microcanonical,fava25designs} with the ideas of unitary designs~\cite{Dankert09unitary} and the theory of free probability \cite{Voiculescu1991}.
The most important object for our work here is the smooth function $\mathcal{O}(\overline{E})$ (we set $\overline{E} \to E$ further on), which was argued in~\cite{Mierzejewski20} to encode the overlap of the Hamiltonian with conserved quantities, including the higher-moments of the Hamiltonian.

In this work, we focus on the ETH in disordered many-body systems
and pose the question of how to perform the disorder average of Eq.~\eqref{eq:ETH} to extract meaningful information about quantum ergodicity of the system. 
We first show that disorder may give rise to anomalous distributions of matrix elements of certain observables, which may be incorrectly interpreted as the breakdown of the ETH.
We then argue that this problem is intimately connected to the way the energy window of the microcanonical ensemble is adjusted to each disorder realization.
We analyze the implications of the different choices of the position of the energy window for the resulting distribution of expectation values of local observables, and we show how properties consistent with the ergodic character of the system are restored.

The paper is organized as follows.
In Sec.~\ref{sec::model}, we briefly describe the peculiar properties of the probability distribution of the site-resolved magnetization for the Heisenberg spin$-1/2$ chain in the presence of the quasiperiodic disorder as encountered in \cite{Aramthottil21}. The site resolved magnetization is defined as  
 $\braket{n\vert \hat{s}_l^z\vert n}\equiv s_l^z$, 
where $\hat s_l^z$ is $z$-projection of spin 1/2 operator at site $l$; with $l$ sampled over site indices and aggregated over sites
when considering distributions.
We proceed to explore the distribution of $s_l^z$ for the Heisenberg spin$-1/2$ chain with different types of onsite disorders. This leads us to a conjecture about the relationship between the distribution of the onsite disorder and the properties of the $s_l^z$ distribution. We show the validity of this conjecture by finding the leading-order approximation for the microcanonical expectation of the site-resolved magnetization in Sec.~\ref{sec::micro}, and conclude that the particular structure of the $s_l^z$ distribution is an artifact of the choice of the energy window, Sec.~\ref{sec::averaging}. Finally, in Sec.~\ref{sec::T_E}, we show that these specific structures have direct consequences for experiments and can be detected in the time dynamics of the $s_l^z$ expectation values.

\begin{figure}
 \begin{center}
\includegraphics[width=\linewidth]{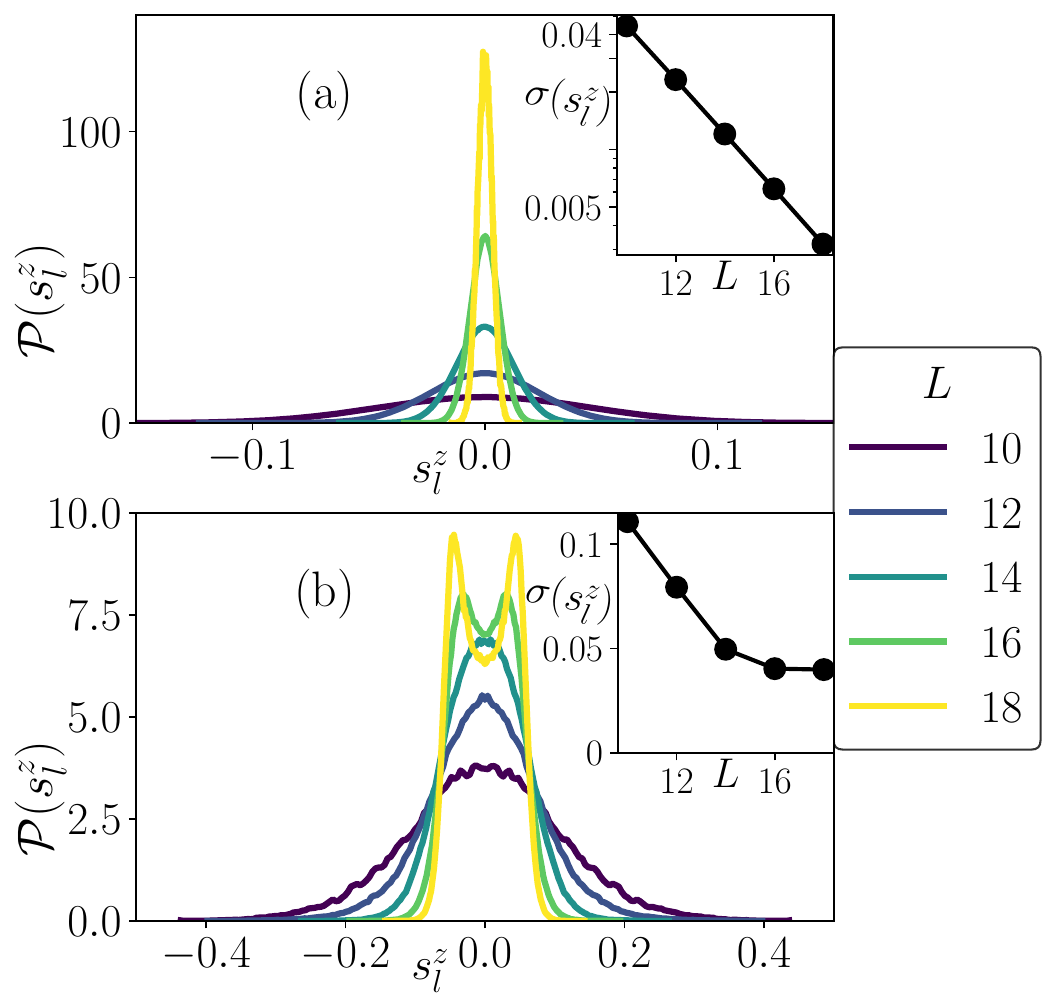}
\caption[]{The probability distribution of site-resolved magnetization, $s_l^z$, for disorder-averaged GOE matrices (a) and the HSC-QP Hamiltonian at $W=0.5$ (b) at different system sizes, $L$. 
For both (a) and (b) panels, $100$ eigenstates are considered at the rescaled energy $\epsilon = 0.5$. The number of disorder realizations is varied, being at least $7500$ for $L\leq 16$ and $60$ for $L=18$ for panel (a), while for HSC-QP case (b) it is $10^4$ disorder realizations for $L\leq 16$ and $2500$ for $L=18$. The insets show the standard deviations of respective distributions, decreasing exponentially with $L$ for GOE (a) and saturating for HSC-QP model.
}
\label{fig:1} 
 \end{center}
\end{figure}

 \begin{figure*}
 \begin{center}
\includegraphics[width=\linewidth]{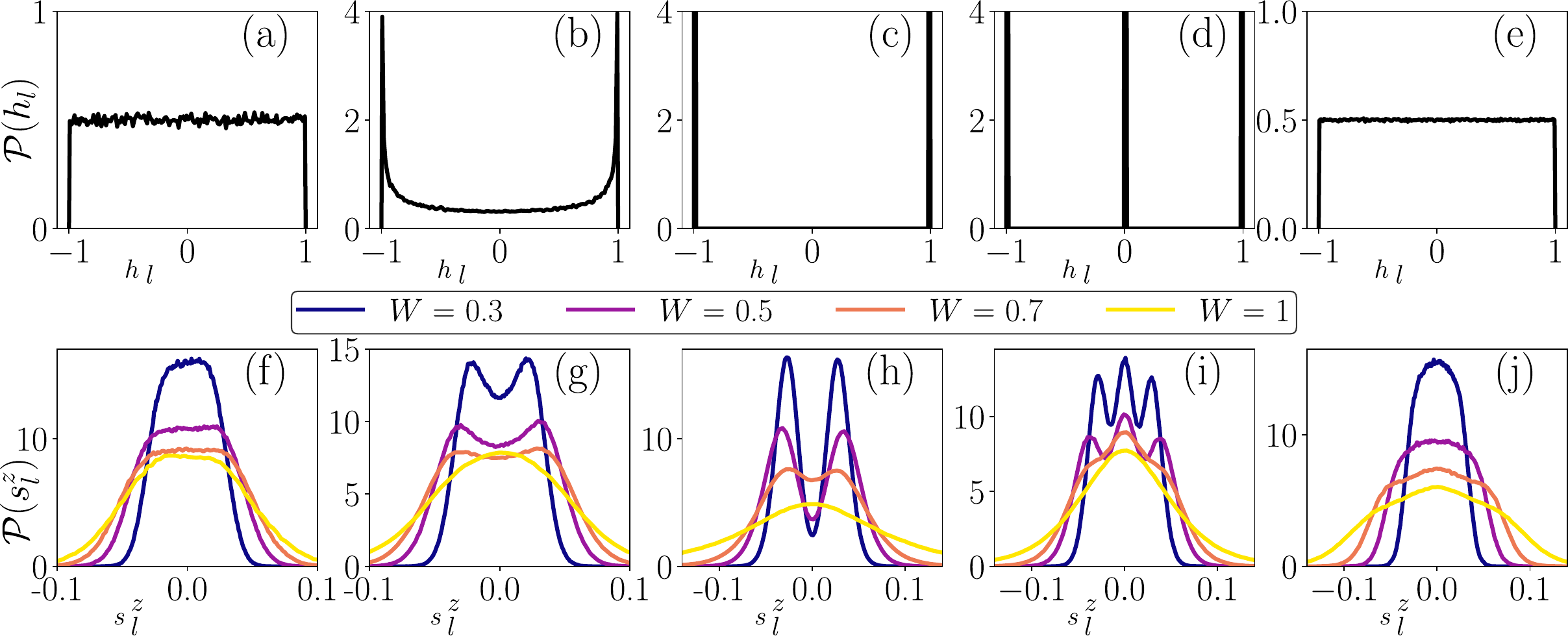}
\caption[Correlated disorder HSC z-spin values]{The distribution of onsite disorder is depicted for various types of disorders, including random (a), cosine (b), binary (c), tertiary (d), and triangle-wave (e). Additionally, the corresponding distributions of site-resolved magnetization are shown in (f-j), considering $100$ states near $\epsilon =0.5$ averaged over at least $200$ disorder realizations for a system of size $L=18$.
}\label{fig:2} 
 \end{center}
\end{figure*}
\section{Unusual distribution for the site-resolved magnetization\label{sec::model}}

We begin with a brief description of the distribution of site-resolved magnetization in a Heisenberg spin-$1/2$ chain (HSC) with quasiperiodic disorder~\cite{Aramthottil21}.
The HSC is a paradigmatic model extensively employed to study ergodicity breaking and the interplay between disorder and quantum many-body dynamics~\cite{Sierant25}. The HSC Hamiltonian with open boundary conditions in the presence of an onsite potential reads
\begin{equation}
\hat{H}_{\mathrm{HSC}} = \sum_{l=1}^{L-1} \big(\hat{s}_l^x\hat{s}_{l+1}^x + \hat{s}_l^y\hat{s}_{l+1}^y + \hat{s}_l^z\hat{s}_{l+1}^z \big) + \sum_{l=1}^L h_l \hat{s}_l^z,
\label{eq:::HSC}
\end{equation}
where $\hat{s}_l^\alpha$ are spin-$1/2$ operators acting on site $l$, with $\alpha \in {x, y, z}$ and $L$ is the system size. Disorder in the model arises from the onsite potential term; initially, we shall consider an example of disorder arising from a quasiperiodic potential (\textbf{QP})~\cite{Iyer13}. The model has a $U(1)$ symmetry leading to magnetization, $M\equiv \sum_{l=1}^L\hat{s}_l^z$, being conserved. We will restrict ourselves to the magnetization-zero sector in this work. The QP introduces a deterministic, incommensurate modulation in the onsite potential, given by:
\begin{equation}
h_l = W\cos(2\pi \beta l + \phi),
\label{eq::QP}
\end{equation}
where $\beta$ is an irrational number, frequently chosen as the inverse golden ratio, ${\beta = (\sqrt{5} - 1)/2}$, ensuring the potential lacks periodicity for typical finite $L$. The phase $\phi$ is drawn randomly from a uniform distribution, $\phi \in [0, 2\pi)$, with each value of $\phi$ corresponding to a distinct disorder realization. At a strong onsite potential, the system is in the non-ergodic, many-body localized (MBL), regime~\cite{Nandkishore15, Alet18,Abanin19, Sierant25} 
which, at system sizes and timescales relevant to experimental and numerical considerations, arises for $W_c\geq 2$ \cite{Aramthottil21, Falcao24}. Interestingly, such a QP disorder appears often in experimental implementations of the disorder \cite{Schreiber15,Bordia16,Bordia17,Luschen17,Luschen18}. In the remainder of this work, however, we focus on the regime of weak disorder, $W \ll W_c$, in which the system is ergodic and possesses level statistics characteristic of the Gaussian Orthogonal Ensemble (GOE) of random matrices~\cite{Mehtabook, Haakebook}.

Analysis of the properties of matrix elements $\bra{n}\hat{\mathcal{O}}\ket{n}$ of the observable $\hat{\mathcal{O}}$ requires fixing an energy window to which the eigenvalues $E_n$ belong. This energy window is commonly rescaled with the system size, $L$, to facilitate comparisons across different system sizes and scales, as detailed in prior works such as \cite{Kim14}. In a more rigorous framework, particularly in the context of studies exploring the MBL phenomenon, a definition of rescaled energy values was proposed in \cite{Luitz15,Luitz16}. This approach provides a 
way of comparing properties of the system at different energies by defining the rescaled energy $\epsilon$ as: 
\begin{equation} \epsilon = \frac{E - E_{\min}}{E_{\max} - E_{\min}}, 
\end{equation} 
where $E$ represents the energy of the state under consideration, while $E_{\min}$ and $E_{\max}$ denote the minimum and maximum 
eigenvalues in the spectrum, respectively. This definition ensures that $\epsilon$ is normalized to the range $[0, 1]$.

 \begin{figure*}
 \begin{center}
\includegraphics[width=\linewidth]{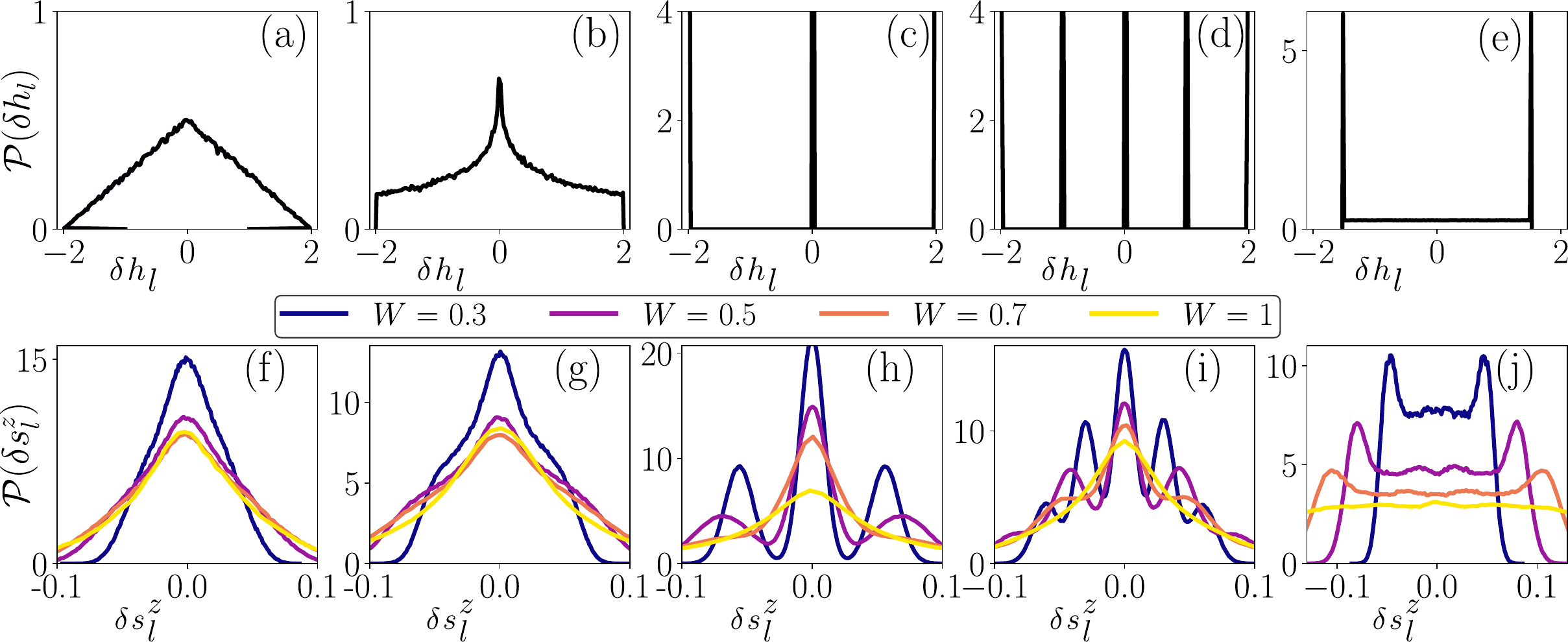}
\caption[Correlated disorder HSC z-spin values]{The distribution of the difference of adjacent onsite disorder, $\delta h_l$, is depicted for various types of disorders, including random (a), cosine (b), binary (c), tertiary (d), and triangle-wave (e). Additionally, the corresponding distributions of differences of adjacent site-resolved magnetization, $\delta s_l^z$, (f-j), considering $100$ states near $\epsilon =0.5$, averaged over at least $200$ disorder realizations for a system of size $L=18$.
}\label{fig:3} 
 \end{center}
\end{figure*}

%%%%%%%%%%%%%%%%%%%%%%

The observable of interest in this work is the site-resolved magnetization, $\hat{\mathcal{O}} \equiv \hat{s}_l^z$. In analogy with a random matrix, one could expect the observable to be Gaussian distributed around its mean value
and with a standard deviation that is exponentially decreasing with system size. We confirm this expectation numerically by taking the Hamiltonian $\hat{H}$ as a matrix from the GOE ~\cite{Mehtabook}. The dimension of $H$ is the same as the Hilbert space dimension of HSC; the diagonal matrix elements of $H$ are taken from the Gaussian distribution $\mathcal{N}(0,1)$ with zero mean and unity variance, while the off-diagonal elements of $H$ are taken from $\mathcal{N}(0,1/{2})$. 
In Figure~\ref{fig:1}(a), we show the probability distribution of site-resolved magnetization considering the expectation values $s^z_l = \braket{n|\hat{s}_l^z|n}$ in eigenstates $\ket{n}$ around the rescaled energy $\epsilon=0.5$. We observe a clear single peak structure with the standard deviation decreasing exponentially with system size.

A different result is obtained for the HSC Hamiltonian with QP disorder for eigenstates around $\epsilon =0.5$, that is, eigenstates in the middle of the spectrum. In this case, the distributions were found~\cite{Aramthottil21} to have a peculiar double-peak structure in the ergodic regime. With increasing system size, the double peak structure was found to be more prominent, as shown in Fig.~\ref{fig:1}(b). 
This result raises the question: Is it a signature of the ETH breakdown, or is it a peculiar property of the disorder distribution?

We note that a similar behavior was found for the difference in the site-resolved magnetization of adjacent sites $\delta s_l^z \equiv s_{l+1}^z-s_l^z$. 
Ref.~\cite{Aramthottil21} hinted at characteristic relationship of the distribution of $s^z_l$ with that of adjacent onsite potential difference, $\delta h_l \equiv h_{l+1}-h_l$, which for QP takes the probability distribution form $\mathcal{P}(\delta h_l)=(2W \pi\sin(\pi \beta))^{-1}(1-(\delta h_l/(2W \sin(\pi \beta)))^2)^{-1/2}$ with $\delta h_l\in[-2W\sin(\pi \beta),2W\sin(\pi \beta)]$ having peaks at $\delta h_l=\pm 2 W\sin{\pi \beta}$. We point out here that the onsite fields $h_l$ also have a probability distribution $\mathcal{P}(h_l)=(W\pi)^{-1}(1-(h_l/W)^2)^{-1/2}$ with peaks at $\pm W$.

Motivated by the intuitions from 
\cite{Aramthottil21} that the distributions of $s_l^z$ or $\delta s_l^z $ have similar characteristics to those of $h_l$ or $\delta h_l$, we here explore the probability distribution of site-resolved magnetization and its differences. In particular, we examine various kinds of disorder and contrast it with the distribution of the onsite disorder and its differences for the following potentials:

\begin{enumerate}
    \item Random disorder ({RD}):
In this case, the onsite potentials $h_l$ are sampled independently from a uniform box distribution:
\begin{equation}
h_l \in [-W, W],
\label{eq::RD}
\end{equation}
where $W$ denotes the disorder strength. The {RD} introduces spatially uncorrelated variations in the potential landscape. This model, at considered system sizes, is in the MBL regime at a sufficiently strong disorder.
In Fig.~\ref{fig:2}(a,f) and Fig.~\ref{fig:3}(a,f), we see that the distribution of $s_l^z$ also has a regime where it is flat, resembling the distribution of $h_l$, while the distribution of $\delta s_l^z$ shares features with the triangle distribution of 
$\delta h_l$.
 \item Cosine disorder: It arises via randomizing the quasiperiodic potential in Eq. \eqref{eq::QP} by choosing the phase $\phi_l\in [0,2\pi)$ independently at each lattice site $l$, \cite{Khemani17}, resulting in
    \begin{equation}
        h_l=W\cos(2\pi \beta l +\phi_l ).
    \end{equation}
The HSC with this disorder type is in the MBL regime for $W\geq 2.5$ \cite{Khemani17}. The cosine disorder as shown in Fig.~\ref{fig:2}(b,g) and Fig.~\ref{fig:3}(b,g) has similar structure of disorder to quasiperiodic one with singular values at the end for $\mathcal{P}(h_l)$. In the ergodic regime of HSC this leads to double peaks appearing in the site-resolved spin expectation values $s^z_l$. At the same time, the distribution of the difference of the on-site potential $\mathcal{P}(\delta h_l)$ differs and shows just a single peak centered around zero. Correspondingly, the distribution of the difference of site-resolved spin expectation values ($\delta s_l^z$) shows a single peak.
    \item Binary disorder: we consider here a distribution with equal probability $h_l=-W$ and $h_l=W$. The model is in the MBL regime for disorder strength $W
    _c\geq1.25$ \cite{Janarek18}. We show in Fig.~\ref{fig:2}(c,h) and Fig.~\ref{fig:3} (c,h) that the distributions of $s^z_l$ in the ergodic regime of this model are characterized by
    strong double peaks resembling the binary distribution of $h_l$. Analogous behavior occurs for $\delta s^z_l$, which has a distribution similar to the distribution of $\delta h_l$, with three peaks, with one in the middle with a prominent amplitude.
    \item Tertiary disorder: we consider a distribution of disorder with $h_l$ selected randomly, with equal probability, to be $h_l\in \{-W,0,W\}$. 
This model is in the MBL regime for $W_c\geq1.25\sqrt{\frac{3}{2}}$ \cite{Janarek18}. In Fig.~\ref{fig:2} (d,i) and Fig.~\ref{fig:3} (d,i), we show the presence of triple peaks in the $s^z_l$ distribution, associated with the tertiary distribution of $h_l$, while $\delta h_l$ has a quinary distribution and correspondingly $\delta s_l^z$ has $5$ peaks with amplitudes decreasing with the distance from $\vert \delta s_l^z\vert =0$.
    \item Triangle-wave quasiperiodic disorder: We consider here a spatially correlated disorder of the form  
\begin{equation}
     h_l=W\left[ 2\left \vert 2 \left(\frac{(2\pi\beta l +\phi)}{2\pi}- \left \lfloor\frac{(2\pi\beta l + \phi)}{2\pi}+\frac{1}{2}\right\rfloor\right)\right\vert-1 \right]\;,
\end{equation}
where $\beta$ is the inverse golden ratio. As shown in Fig.~\ref{fig:2} (e,j) and Fig.~\ref{fig:3} (e,j), this disorder has the peculiar property of $h_l$ being distributed uniformly, resulting in $s^z_l$ admitting a flat-top distribution form similar to RD, while the distribution of $\delta h_l$ has singularities at the edges of the distribution corresponding to double peaks for $\delta s_z^l$.
\end{enumerate}

The numerical analyses from Figs.~\ref{fig:2} and~\ref{fig:3} confirm the inference of \cite{Aramthottil21}: for each type of disorder, the site-resolved magnetization (or its difference between neighboring sites) has a distribution that closely resembles the features of the distribution of on-site disorder (or its differences between neighboring sites). These features are inconsistent with the distribution of expectation values of local observables in highly excited eigenstates based on random matrix theory, cf. Fig.~\ref{fig:1}~(a), and hence may be interpreted as inconsistent with the ergodicity of the system. The remainder of this manuscript is devoted to understanding the origin and the implications of these \textit{false} signatures of non-ergodicity in the ergodic regime of disordered many-body systems.

\section{Microcanonical expectation for site-resolved magnetization\label{sec::micro}}
In this section, we argue that the apparent signatures of non-ergodicity in the disorder-averaged distributions of matrix elements of local observables---shown in the bottom rows of Fig.~\ref{fig:2} and Fig.~\ref{fig:3}---originate from the interplay between the energy dependence of $\mathcal{O}(E)$ in the ETH ansatz~\eqref{eq:ETH} and the choice of energy window ($\epsilon = 0.5$) used in the analysis.

To proceed more formally, following~\cite{Mierzejewski20} we define the Hilbert-Schmidt product in operator space as 
$\braket{\hat{A},\hat{B}}\equiv \frac{1}{{D}}\mathrm{Tr}[\hat{A}^{\dagger}\hat{B} ] $, where \({D}\) is the Hilbert space dimension, and the norm as 
 $|| \hat{A} || = \sqrt{\braket{\hat{A},\hat{A}}}$. The Hamiltonian of an ergodic 
many-body system is denoted by $\hat{H}$, and $\hat{\mathcal{O}}$ represents an 
arbitrary few-body observable. The operator space can be decomposed into two orthogonal 
subspaces: one spanned by the powers of the Hamiltonian,  $\hat{H}^k$ (with $k \geq 0$), and an orthogonal complement. Hence, the observable 
\(\hat{\mathcal{O}}\) can be decomposed as
\begin{equation}
\hat{\mathcal{O}} = \hat{\mathcal{O}}_{\|}+\hat{\mathcal{O}}_{\perp},
\label{eq:O1}
\end{equation}
where $\hat{\mathcal{O}}_{\|}$ belongs to the subspace spanned by the moments of the Hamiltonian $\hat{H}^k$, while $\hat{\mathcal{O}}_{\perp}$ lies in the orthogonal complement of this subspace. 
Assuming that $\ell$ is the smallest positive integer such that $\hat{H}^{\ell+1}$ is linearly dependent on $\hat{H}^{0}, \hat{H}, \hat{H}^2,\ldots, \hat{H}^{\ell}$, the parallel component of the observable of interest can be uniquely expressed as 
\begin{equation}
\hat{\mathcal{O}}_{\|}=\sum_{k=0}^{\ell} \gamma_k \hat{H}^k.
\label{eq:O2}
\end{equation}
Substituting Eq.~\eqref{eq:O2} into Eq.~\eqref{eq:O1} and evaluating the expectation value 
in the eigenstate $\ket{n}$ of the Hamiltonian $\hat{H}$, we obtain
\begin{equation}
\braket{n|\hat{\mathcal{O}}|n}=\sum_{k=0}^{\ell} \gamma_k E_n^k +  \braket{n|\hat{\mathcal{O}}_{\perp}|n}.
\label{eq:O3}
\end{equation}
To establish a connection between the diagonal matrix element $ \braket{n|\hat{\mathcal{O}}|n}$ and the smooth function $\mathcal{O}(E)$ in the ETH ansatz~\eqref{eq:ETH},  it is necessary to understand the behavior of $\braket{n|\hat{\mathcal{O}}_{\perp}|n}$.

 \begin{figure}
 \begin{center}
\includegraphics[width=\linewidth]{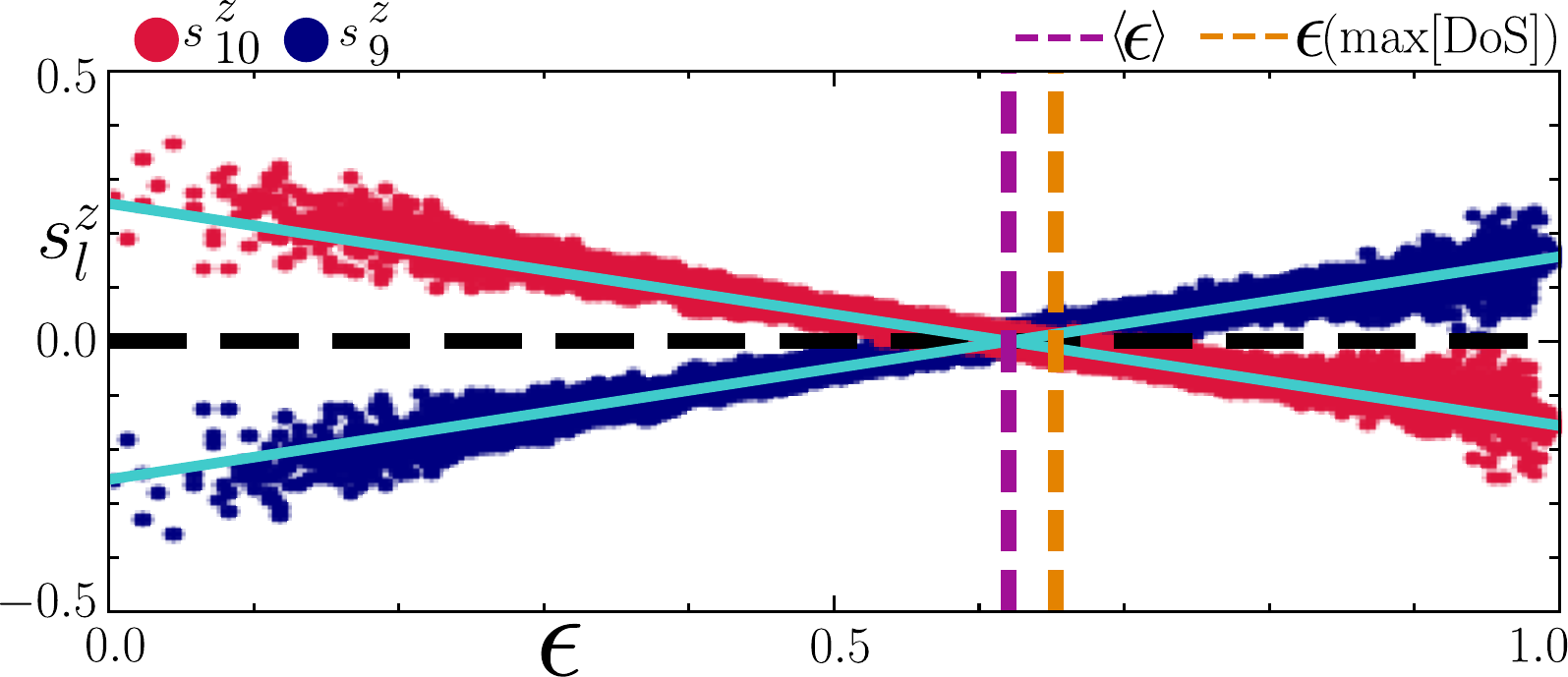}
\caption[QP HSC z-spin values]{
The site-resolved magnetization, $s_l^z$, for a single disorder realization of the HSC QP model at different rescaled energies $\epsilon$ for $L=18$ and $W=0.5$. The dark blue and red scatter plots correspond to sites $l=9$ and $l=10$. The dashed, vertical lines indicate the rescaled energy values that give the maximum value of $\mathrm{DoS}$ and the average rescaled energy $\braket{\epsilon}$. The bold blue lines illustrate the analytical curves obtained by expanding the microcanonical expectation value, $\mathcal{O}(\epsilon)$, to the first order in the Hamiltonian's moment. 
}\label{fig:4} 
 \end{center}
\end{figure}

By definition, since $\hat{\mathcal{O}}_{\|}$ is a linear combination of the powers of the Hamiltonian $\hat{H}$, we have $[\hat{H}, \hat{\mathcal{O}}_{\|}] = 0$. Therefore, the operator $\hat{\mathcal{O}}_{\perp}$ encodes the components of the observable $\hat{\mathcal{O}}$ that do not commute with the Hamiltonian. In particular, when $||[\hat{H}, \hat{\mathcal{O}}]|| = O(1)$, the norm of $\hat{\mathcal{O}}_{\perp}$ is also of order unity, $||\hat{\mathcal{O}}_{\perp}|| = O(1)$.
To bound \(\braket{n|\hat{\mathcal{O}}_{\perp}|n}\) in Eq.~\eqref{eq:O3}, we invoke properties of ergodic systems. While $\hat{\mathcal{O}}_{\perp}$ is a nontrivial operator, it bears no special relationship to the eigenvectors $\ket{n}$ of the Hamiltonian. If the system is ergodic, $\braket{n|\hat{\mathcal{O}}_{\perp}|n}$ can be modeled as a diagonal element of a random matrix scaling as  $\braket{n|\hat{\mathcal{O}}_{\perp}|n} = O(1/\sqrt{D})$ and, therefore, exponentially with the system size, $L$. In this case, Eq.~\eqref{eq:O3} implies that $\braket{n|\hat{\mathcal{O}}|n}$ becomes, up to \(O(e^{-L})\) corrections, a smooth function of energy, leading to
\begin{equation}
\mathcal{O}(E)=\sum_{k=0}^{\ell} \gamma_k E^k.
\label{eq:O4}
\end{equation}
In passing, we note that the influence of integrals of motion beyond the moments of the Hamiltonian, $\hat{H}^k$, can be systematically taken into account, as shown in \cite{Mierzejewski20}.

\begin{figure}
 \begin{center}
\includegraphics[width=\linewidth]{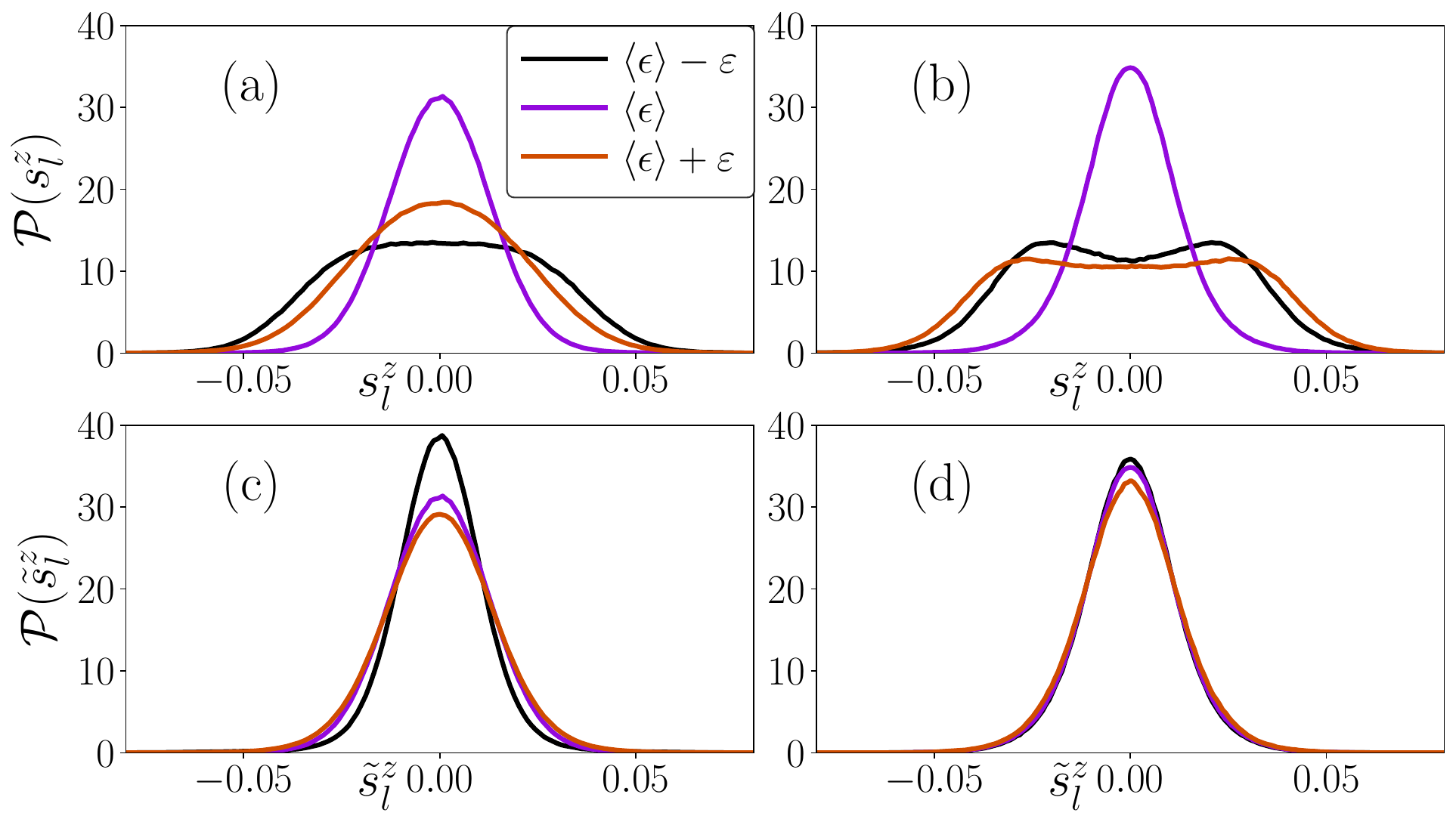}
\caption[]{(a,b) The probability distribution for the site-resolved magnetization, $s_l^z$, and (c,d) the site-resolved magnetization after removing the approximate microcanonical expectation, $\tilde{s}_l^z$.
Results are shown, respectively, for the HSC model with Random (a,c) and QP (b,d) disorder. For all panels $L=18$, $W=0.5$, with $100$ eigenstates in intervals around $\braket{\epsilon}$ and  $\braket{\epsilon}\pm \varepsilon$ for at least $1300$ disorder realizations.}
\label{fig:mic_rem} 
 \end{center}
\end{figure}

Let us now specialize to the observable of interest, i.e., the site-resolved magnetization, $\hat{s}^z_l$, and as the Hamiltonian consider $\hat{H}_{\mathrm{HSC}}$. From Eq.~\eqref{eq:O4}, we observe that understanding of the microcanonical expectation value $s^z_l(E)$ requires evaluation of the coefficients $\gamma_k$. Moreover, for a system with near-Gaussian density of states (DoS) with a variance proprtional to the system size (see Appendix \ref{ap:dens}) and an operator with norm independent of the system size, it was observed~\cite{Mierzejewski20} that taking only the first $p$ terms in the expansion~\eqref{eq:O2}, incurs error decreasing with system size $L$ as $O(1/L^{p-1})$. For this reason, we consider only the first two terms in the expansion $\hat{s}^z_{l,\|} = \gamma_0 \hat{H}_{\mathrm{HSC}}^0 + \gamma_1 \hat{H}_{\mathrm{HSC}}^1$, resulting in 
\begin{equation}
\hat{s}^z_l = \gamma_0 \mathbb{I} + \gamma_1 \hat{H}_{\mathrm{HSC}}+\hat{s}^z_{l,\perp}.
\label{eq:O5}
\end{equation}
Calculating the Hilbert-Schmidt product $\braket{ \mathbb{I},\hat{s}^z_l }$ in \eqref{eq:O5}, we find that  
\begin{equation}
\frac{1}{D} \mathrm{Tr}(\hat{s}^z_l)= \gamma_0+ \gamma_1\frac{1}{D} \mathrm{Tr}(\hat{H}_{\mathrm{HSC}}),
\label{eq:O6}
\end{equation}
while, by computing  $\braket{ H_{\mathrm{HSC}},\hat{s}^z_l }$ in \eqref{eq:O5}, we obtain that 
\begin{equation}
 \frac{1}{D} \mathrm{Tr} (\hat{H}_{\mathrm{HSC}} \hat{s}^z_l ) = \gamma_0 \frac{1}{D}\mathrm{Tr}(\hat{H}_{\mathrm{HSC}}) + \gamma_1 \frac{1}{D} \mathrm{Tr}( \hat{H}_{\mathrm{HSC}}^2).
\label{eq:O7}
\end{equation}
In both steps, we used the fact that $\hat{s}^z_{l,\perp}$ is orthogonal to $\hat{H}^k_{\mathrm{HSC}}$ for any integer $k\geq 0 $.
Directly computing $\mathrm{Tr}(s^z_l) =0$, $\mathrm{Tr}(\hat{H}_{\mathrm{HSC}})/{D}=-1/4$, $\mathrm{Tr}(\hat{H}_{\mathrm{HSC}}\hat{s}_l^z)/{D}=1/4h_l-1/(4(L-1))\sum_{l\neq i}h_i$  and $\mathrm{Tr}(\hat{H}_{\mathrm{HSC
}}^2)/{D}=(3L^2-3L-1)/(16(L-1))+1/4\sum_{i=1}^Lh_i^2-1/(4(L-1))\sum_{l\neq j}h_lh_j$, we use Eq.~\eqref{eq:O6} and~\eqref{eq:O7} to determine the values of $\gamma_0$ and $\gamma_1$, which, through \eqref{eq:O4}, imply that 
\begin{equation}
s_j^z(E)\propto \frac{E+1/4}{3L/4+\sum_ih_i^2}h_j=\frac{E-\braket{E}}{3L/4+\sum_ih_i^2}h_j \;,
\label{eq:szj}
\end{equation}
up to $O(1/L)$ corrections. Above, $\braket{E} = \frac{1}{D} \mathrm{Tr}(\hat{H}_{\mathrm{HSC}}) = -\frac{1}{4}$.
Further details of the 
derivations may be found in Appendix~\ref{ap:anal} while the near-Gaussian assumption on the density of states is justified in Appendix~\ref{ap:dens}.

\begin{figure}
 \begin{center}
\includegraphics[width=\linewidth]{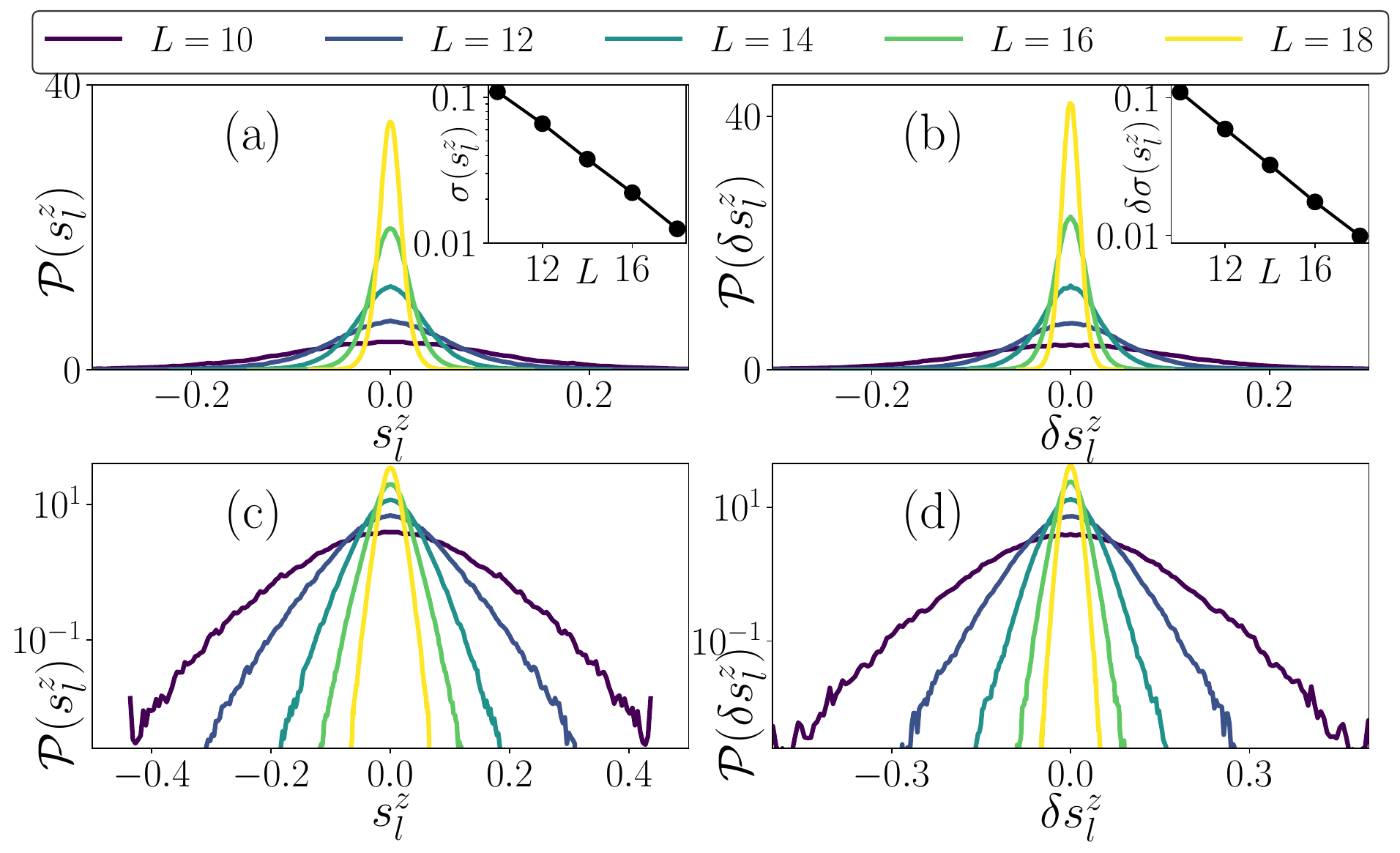}
\caption[]{The probability distribution of site-resolved magnetization, $s_l^z$, (a) and the difference of site-resolved magnetization, $\delta s_l^z$, (b) for disorder-averaged HSC-QP Hamiltonian. The insets for (a) and (b) show the standard deviation with system sizes for $s_l^z$ and $\delta s_l^z$, respectively. Panels (c) and (d) show the same curves as in (a) and (b) but in a logarithmic vertical scale.  The distributions are for $W=0.5$ considering $100$ eigenstates at the average energy $\braket{\epsilon}$ with at least $4500$ disorder realizations for different system sizes, $L$.
}
\label{fig:9} 
 \end{center}
\end{figure}
The derived linear relationship between the site-resolved magnetization, $s_j^z(E)$, and the energy 
accurately reproduces the behavior of the diagonal matrix element $\braket{n|\hat{s}^z_l|n}$ in the Hamiltonian's eigenstates. 
This is illustrated in Fig.~\ref{fig:4} which shows that $\braket{n|\hat{s}^z_l|n}$ concentrate, up to fluctuations suppressed exponentially with system size $L$, around the curves $s_j^z(E)$ at all values of the rescaled energy $\epsilon$ as ascertained in Fig.~\ref{fig:9}. Fitting $\braket{n|\hat{s}^z_l|n}$ as a function of energy, we confirm that the resulting linear fit reproduces the parameters of \eqref{eq:szj} down to machine precision. In particular, we observe that the zero of $s_l^z(E)$ occurs at the rescaled energy $\braket{\epsilon}$ corresponding to $\braket{E}$. 

%%%%%%%%%%%%%%%%%%%%%%%%%%%%%%%%%%%%%%%%%%%%%%%%%%%%%%%%%%%%%%%%%%%%%%%%%%%%%%%

\section{Artifacts in disorder averaged observables\label{sec::averaging} }
The simple approximation for the microcanonical
expectation of site-resolved magnetization, \eqref{eq:szj},
establishes a framework to capture the anomalous probability distribution of ${s}^l_z$ when disorder-averaged properties of the system are considered.

The approximation implies that for a given disorder strength, $s_l^z$ is proportional to the local disorder field $h_l$ and linearly varies with the energy from the point $E=\braket{E}$. Consequently, if one considers the eigenstates around an energy significantly distant from $\braket{E}$ for different disorder realizations, the $s_l^z$ values will be confined within a width restricted by the energy distance from $\braket{E}$, and they will take linearly dependent values proportional to $h_l$ within this width. This scenario is illustrated in Fig.~\ref{fig:mic_rem} where we compare the distributions of $s_l^z$ obtained for different energy intervals, one of them being centered at $\braket{\epsilon}\approx 0.62$ and two other at $\braket{\epsilon}\pm \varepsilon$ with $\varepsilon\approx 0.076$. Already for random uniform disorder, the obtained distributions significantly differ, compare  Fig.~\ref{fig:mic_rem}(a). For QP disorder, the difference is more spectacular, and a double peak structure appears. 

Consider now ``corrected'' observables $\tilde{s}_l^z=s_l^z-s_l^z({E})$, where $s_l^z(E)$ is the approximation for the microcanonical expectation found by explicitly solving for the first two coefficients in~\eqref{eq:O4}. Such a compensation results in a single peak, Gaussian-like 
distributions for both random uniform and QP disorders, as illustrated in Fig.~\ref{fig:mic_rem}(c,d). We have verified that the residual differences  (absent, by the way,  in the QP case) can be attributed to finite-size (by a comparison with lower system size results) and low-order truncation effects.

Let us concentrate on
eigenstates around $\braket{\epsilon}$ where no energy correction to magnetization is necessary and analyze the influence of the system size on the probability distributions for both $s_l^z$ and $\delta s_l^z$. Both distributions significantly narrow with the 
system size, with the associated standard deviation decreasing exponentially with increasing $L$, as shown in insets in Fig.~\ref{fig:9}. The distributions are, to a good approximation, Gaussians for sufficiently large $L$ as apparent from the lower panels plotted in the logarithmic vertical scale.
This indicates that the corrections to $s_l^z$ coming from the energy dependence are exponentially decreasing with $L$, consistently with the expectations based on the random matrix theory, recall~Fig.~\ref{fig:1}(a).

\begin{figure}
 \begin{center}
\includegraphics[width=\linewidth]{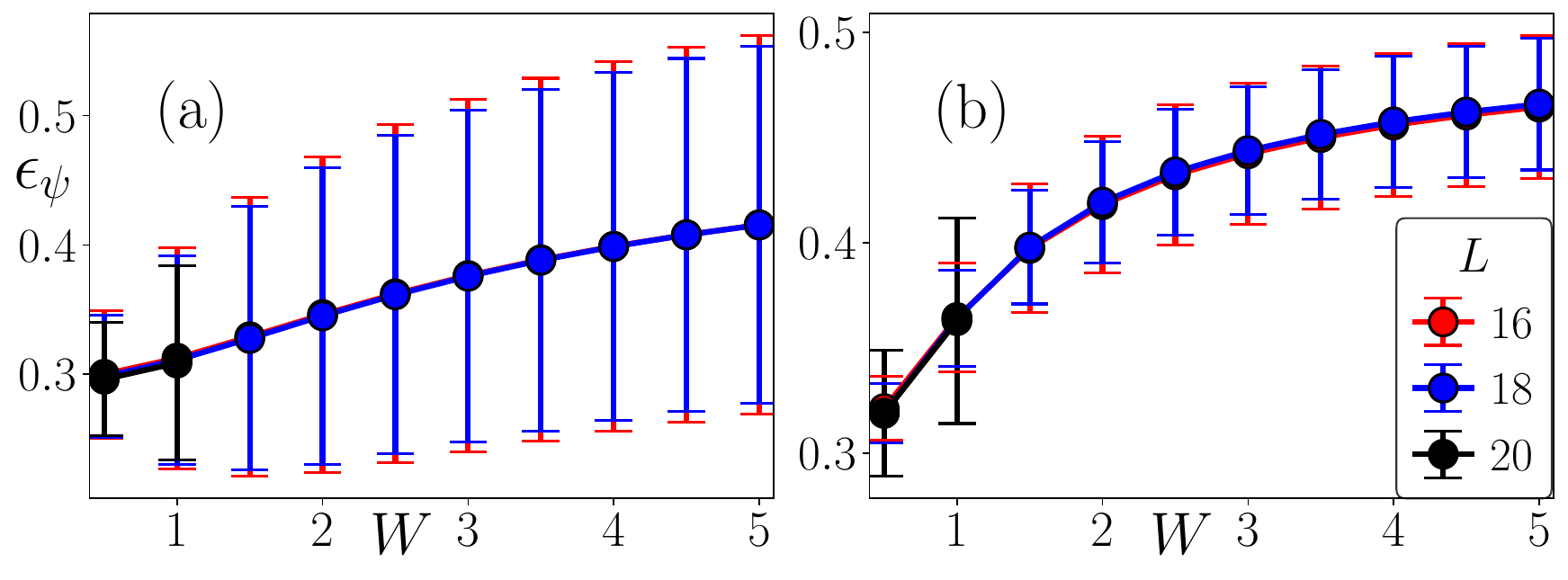}
\caption[]{The disorder averaged rescaled energy expectation value for the N\'eel state, $\epsilon_{\psi}$, and the associated standard deviation with disorder realization (shown here by the error bars ) varying with disorder strength for different system sizes. For the HSC Hamiltonian with RD (a) and QP disorder (b). Both (a) and (b) consider $2500$ disorder realizations.
}\label{fig:7} 
 \end{center}
\end{figure}

Within this framework, one can explain the anomalous behaviour in Fig.~\ref{fig:1}(b), as well as similar observations in \cite{Hopjan20,Hopjan21,Laflorencie20}. The rescaled energy considered in that work, fixed as $\epsilon =0.5$, significantly deviates from the value of the rescaled energy $\braket{\epsilon}\approx 0.62$ corresponding to $\braket{E}=-1/4$.
Note also that $\braket{\epsilon}$ fluctuates between disorder realizations.  In effect
one observes non-trivial distributions of the on-site magnetization as in Fig.~\ref{fig:2}, where the eigenstates belong to the energy window determined by $\epsilon =0.5$ sufficiently far from $\braket{\epsilon}$ and also fluctuate from one disorder realization to another. 
The same mechanism can also be extended to linear combinations of site-resolved magnetizations, explaining the characteristics of differences in site-resolved magnetization shown in Fig.~\ref{fig:3}.

The interested reader may be puzzled by the dependence of the observed anomalies on disorder amplitude $W$. Firstly, let us stress that such a behavior may be considered as anomalous when observed in the otherwise ergodic regime. For large disorder, close to the crossover to many-body localization, the nonergodic behavior is expected. The small disorder details are discussed in Appendix~\ref{ap:new}.

\begin{figure}
 \begin{center}
\includegraphics[width=\linewidth]{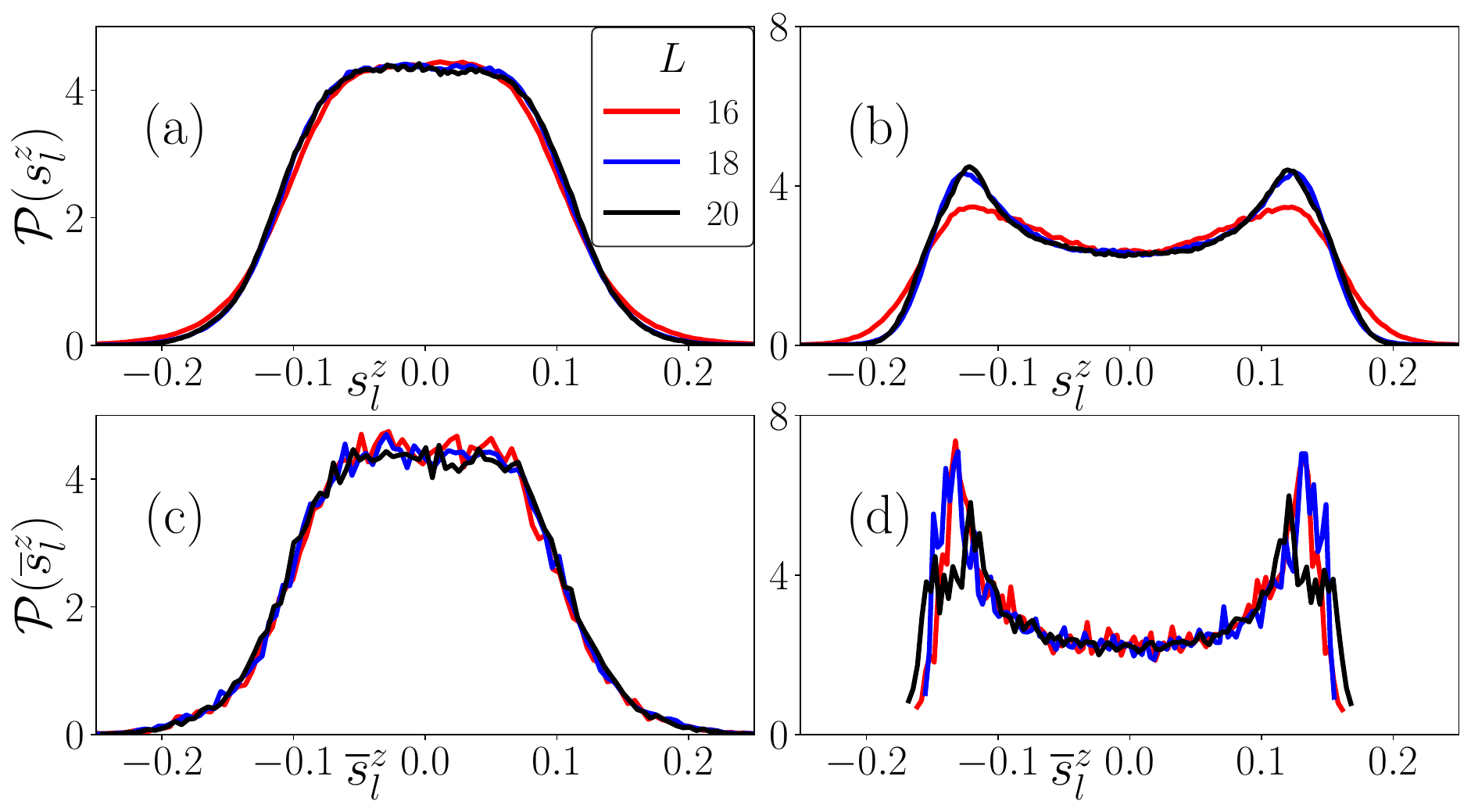}
\caption[]{The probability distribution of site-resolved magnetization of eigenstates, $\mathcal{P}(s_l^z)$, for different system sizes around energies $\braket{\psi\vert \hat{H}\vert \psi}$, the expectation energy of the  N\'eel state for both the RD (a) and QP (b) case. Both (a) and (b) consider $100$ eigenstates for at least $1000$ disorder realizations with $W=0.5$. The bottom row provides a comparison with the distributions for site-resolved magnetization at long times obtained from dynamics of the initial  N\'eel state, $\mathcal{P}(\overline{s}_l^z)$ again for both RD (c) and QP (d) disorder cases. Both (c) and (d) consider $2500$ disorder realizations.}
\label{fig:6} 
 \end{center}
\end{figure}
\section{Effects on time-dynamics\label{sec::T_E}}

In this section, we examine the experimental consequences of the anomalous distributions of $s_l^z$ in eigenstates. {We consider a system with Hamiltonian $\hat{H}$, initialized in state $\ket{\phi}$ and denote the time-evolved state by $\ket{\phi(t)} = e^{-i\hat{H}t }\ket{\phi}$. Long time average $\overline{\mathcal{O}}$ of expectation value of an observable $\hat{O}$ is given as}
\begin{align}
 \nonumber
    \overline{\mathcal{O}} &{\equiv} \lim_{T\rightarrow \infty}{\frac{1}{T}}\int_{0}^T dt \braket{\phi(t) \vert \hat{\mathcal{O}}\vert  \phi(t)}=
   \\ &{=\lim_{T\rightarrow \infty} \frac{1}{T}\int_{0}^T dt \, e^{i(E_m-E_n)t}\sum_{m,n} \braket{m| \hat{\mathcal{O}}|n}c_m^* c_n,} 
    \label{eq:lala}
\end{align}
where $c_n = \braket{n|\phi}$. Performing the time integral, we note that only the diagonal terms, $m=n$, yield a non-vanishing contribution, showing that $\overline{\mathcal{O}}$ is given by the expectation value in the \textit{diagonal ensemble,}
\begin{equation}
    \overline{\mathcal{O}} = \sum_{n} \braket{n| \hat{\mathcal{O}}|n} |c_n|^2,
    \label{eq:diagEns}
\end{equation}
defined by the initial state $\ket{\phi}$. When the $\hat{H}$ is ergodic and local, and the initial state is a product state, the energy fluctuations are subextensive~\cite{Dalessio16}, $\delta E_\phi/E_\phi \sim 1/\sqrt{L}$, where $E_\phi = \braket{\phi|\hat{H}|\phi}$ and $\delta E_\phi=\sqrt{\braket{\phi\vert \hat{H}^2 \vert \phi}-E_\phi^2}$. In that case, the diagonal ensemble expectation value, \eqref{eq:diagEns}, may be close to the microcanonical average $\mathcal{O}(E_{\phi})$, \eqref{eq:ETH}, at energy $E_{\phi}$. 
In an ergodic disordered system, the mechanism leading to the \textit{false} signatures of non-ergodicity in the distributions of expectation values of local observables in eigenstates may lead to a {non-trivial spread of the long-time averages, Eq.~\eqref{eq:lala}, of the local observables}. In the following, we {analyze this effect} on the example of disordered HSC, showing how to assess the ergodicity of the system correctly. 

We fix the initial quench state as the N\'eel state 
$\ket{\psi}\equiv \ket{\uparrow\downarrow \cdots \uparrow\downarrow}$ and follow the time evolution for individual disorder realizations using Chebyshev propagation scheme \cite{Fehske08} - see Appendix~\ref{ap: time}. For disorder strength $W=0.5$, the average rescaled energy of $\ket{\psi}$ in the HSC model with both the RD and QP disorder is $\braket{\psi\vert \hat{H}_{\mathrm{HSC}}\vert \psi}\equiv $ $\epsilon_\psi \approx 0.3$, c.f. Fig. \ref{fig:7}. The latter value is significantly distant from the rescaled energy $\braket{\epsilon}\approx 0.62$ corresponding to $\braket{E}=-1/4$. At the considered system size, the expectation value $\hat{s}_l^z$ enters a stationary regime already at time-scales $O(100)$. Hence, as an approximation for the infinite time average in Eq.~\eqref{eq:lala} for the operator $\hat{s}^z_l = \hat{\mathcal{O}}$, we consider  its long time average as 
\begin{equation}
\overline{s}_l^z \equiv \frac{1}{t_2-t_1} \int_{t_1}^{t_2}\braket{\psi\vert \hat{s}^z_l(t) \vert \psi }dt.
\label{eq:timeave}
\end{equation}
where we take $t_1=200$ and $t_2=2000$. For the considered HSC, and for the operator $\hat{\mathcal{O}} = \hat{s}^z_l$, the long time average approximates accurately (for this choice of $t_1$, $t_2$ up to errors of the order $O(10^{-4})$) the diagonal ensemble expectation value given by~\eqref{eq:diagEns}. Hence, in the following, we are using the diagonal ensemble and the long-time average of the $\hat{s}^z_l$ operator interchangeably. We show in Fig.~\ref{fig:6} that the probability distribution obtained from the long-time average of $\overline{s}_l^z$ over disorder realizations has similar characteristics to that found from eigenstates at energies corresponding to the initial energy of the  N\'eel state for each disorder realization. In particular, for the HSC-QP model, $\mathcal{P}(\overline{s}_l^z)$ reveals a double peak structure, with similar widths of the side peaks for $L=16,18$  while being broader for $L=20$, which reflects the standard deviations in $\epsilon_\psi$ (see Fig. \ref{fig:7}). With increasing system size the  orthogonal fluctuation to the microcanonical expectation, $\braket{n\vert \hat{O}_{\perp}\vert n}$, exponentially decreases leading to better approximations of the peaks in  $\mathcal{P}(\overline{s}_l^z)$ by the distribution $\mathcal{P}(s_l^z)$. 

\begin{figure}
 \begin{center}
\includegraphics[width=0.8\linewidth]{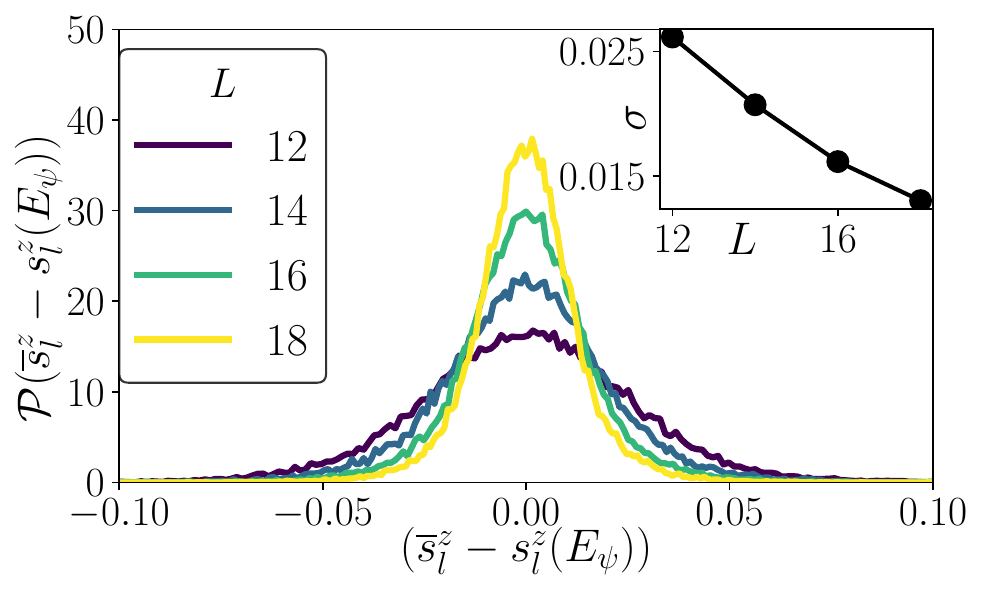}
\caption[]{The probability distribution of long time average of the site-resolved magnetization, $\overline{s}_l^z$, \eqref{eq:timeave}, corrected by the microcanonical expectation value, $s_l^z(E_{{\psi}})$, in the  N\'eel state, $\mathcal{P}(\overline{s}_l^z-s_l^z(E_{{\psi}}))$, for the HSC Hamiltonian with RD for different system sizes. The microcanonical expectation is calculated as an average over $10$ eigenstates, and the probability distributions are found by considering $2500$ disorder realizations.
}\label{fig:mic_rem_TD} 
 \end{center}
\end{figure}

The distributions of $\overline{s}^z_l$, obtained from the simulated time dynamics for different sites $l$ and disorder realizations, do not follow a naive expectation of their narrow concentration around the mean value equal to zero for disordered ergodic systems.
As shown in Fig.~\ref{fig:mic_rem_TD}, these distributions are centered around the microcanonical ensemble expectation value, $s^z_l(E_{\psi})$ calculated at the energy $E_{\psi}$ of the initial N\'{e}el state, up to corrections decreasing with system size $L$. This microcanonical ensemble expectation value is estimated as an average of $s^z_l$ over $N$ eigenstates with eigenenergies close to
a desired energy $s^z_l(E)=1/N\sum_n \braket {n \vert \hat{s}^z_l \vert n} $.

\begin{figure}
\begin{center}
\includegraphics[width=\linewidth]{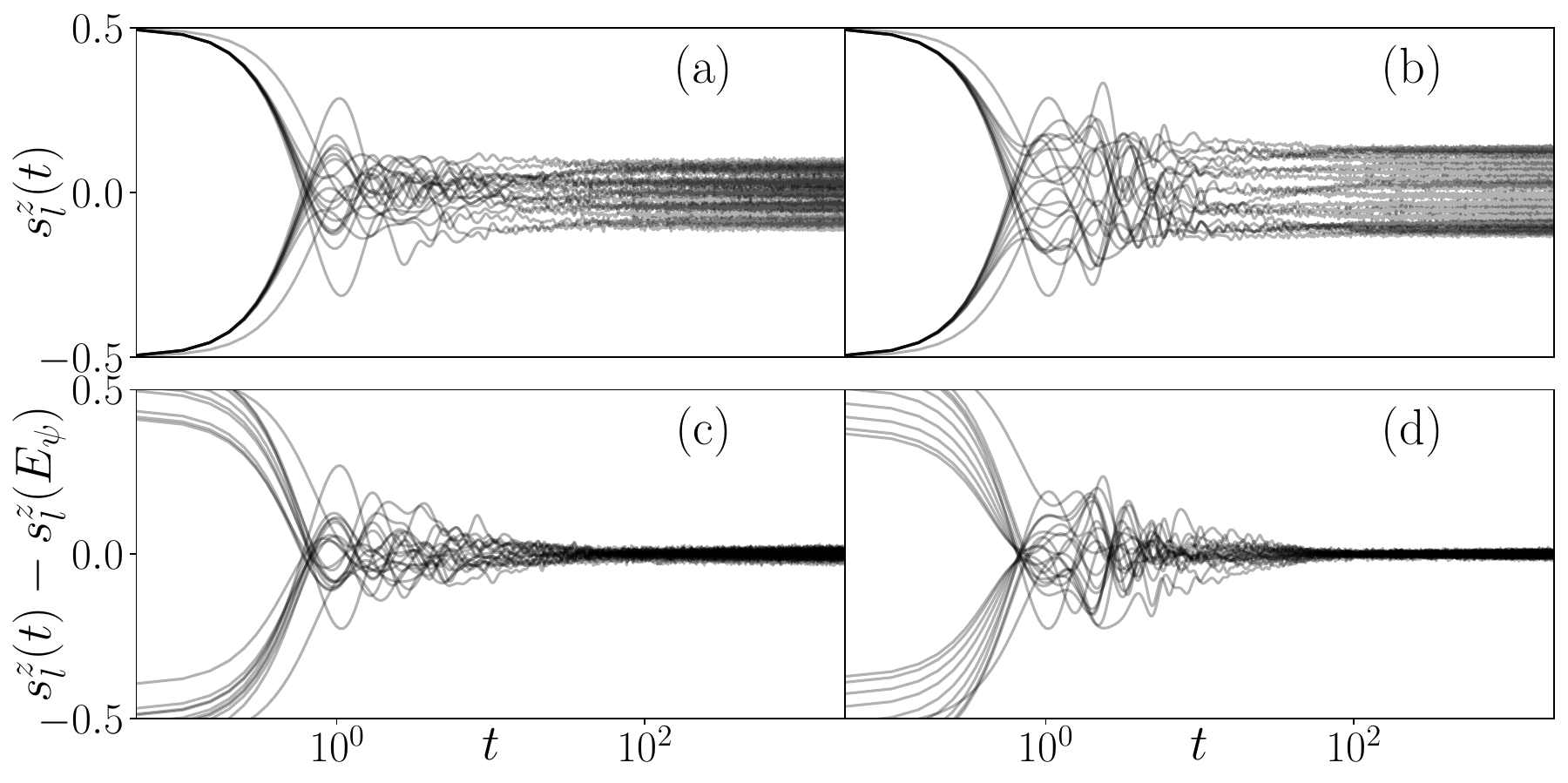}
\caption[]{The time dynamics of the site-resolved magnetizations, $s_l^z(t)=\braket{\psi\vert \hat{s}_l^z(t)\vert\psi}$, for  the N\'eel initial evolved state  with the HSC Hamiltonian and RD, (a) or QP (b) disorders for a single disorder realization with $W=0.5$ and the system size $L=20$. Each black curve represents a different site magnetization  $s_l^z(t)$. The bottom row shows the same dynamics when the microcanonical expectation for each spin, ${s}_l^z(E_{\psi})$ (obtained using 100 eigenstates)  is subtracted, $\tilde{s}_l^z \equiv s_l^z(t)-{s}_l^z(E_{\psi})$.
}\label{fig:sing_dis} 
\end{center}
\end{figure}

Similarly to the case of eigenstates considered earlier, the significant spread of the long-time averages, $\overline{s}^z_l$, may lead to \textit{false} conclusions about the breakdown of ergodicity. A remedy for this artifact may be, however, readily constructed, even at the single disorder realization level. Instead of analyzing the long-time dynamics for individual spin, $s^z_l(t)= \braket{\psi\vert \hat{s}_l^z(t)\vert\psi}$, one may subtract from it the microcanonical expectation.
The result of such a procedure can be seen in
Fig.~\ref{fig:sing_dis}. The top row presents different $s^z_l(t)$ for a single disorder realization, showing the residual spread of the saturation values, $\overline{s}^z_l$, present even at long times.
When $\tilde s^z_l(t)= s^z_l(t)-s^z_l(E_{\psi})$  are plotted, the spread dramatically narrows down, and $\tilde s^z_l(t)$ fluctuates, at long times, around zero mean, as one could expect for the ergodic dynamics.

\section{Conclusion}

In this study, we have highlighted the subtle issues related to self-averaging in disorder-averaged observables, focusing on the experimentally relevant probe of site-resolved magnetization in spin-$1/2$ chains. Taking as an example the paradigmatic Heisenberg spin-$1/2$ chain, we systematically investigated various types of onsite disorder in the ergodic regime. Our results show that the nontrivial structures observed in the probability distributions of matrix elements of local observables arise as a direct consequence of the energy dependence of the microcanonical expectation value, which varies for each specific disorder realization. Importantly, we demonstrated that the expected Gaussian behavior predicted by random matrix theory can be recovered either by selecting eigenstates near the infinite-temperature energy window or by explicitly removing the mean microcanonical contribution. This work clarifies how misleading signatures of non-ergodic behavior can emerge and provides practical guidance for correctly interpreting numerical and experimental results in disordered quantum systems.

Furthermore, we have shown that these microcanonical artifacts persist in experimentally relevant long-time dynamics and can lead to false signatures of non-self-averaging behavior if not properly accounted for. Our results highlight the necessity of carefully distinguishing true ergodicity-breaking effects from energy-window artifacts when interpreting both numerical and experimental observations.

{\bf Data availability statement}
All data that support the findings of this study are available at  RODBUK UJ repository and available at  https://doi.org/10.57903/UJ/BCYX1B .

\acknowledgments
A.S.A., P.S., and J.Z. would like to acknowledge insightful discussions with Titas Chanda in the initial stages of this work. We thank also Konrad Pawlik for a careful reading of the manuscript. The work of  A.S.A. and J.Z. was partially funded by the National Science Centre, Poland, project 2021/03/Y/ST2/00186 within the QuantERA II Programme (DYNAMITE) that has received funding from the European Union Horizon 2020 research and innovation programme under Grant agreement No 101017733 and by the National Science Centre, Poland, project  2021/43/I/ST3/01142 -- OPUS call within the WEAVE programme. 
A.E.K. acknowledges the support of the National Science Centre, Poland, via Project No. 2021/42/A/ST2/00017.
P.S. acknowledges fellowship within the “Generación D” initiative, Red.es, Ministerio para la Transformación Digital y de la Función Pública, for talent attraction (C005/24-ED CV1), funded by the European Union NextGenerationEU funds, through PRTR.
L.V. acknowledges support from the Slovenian Research and Innovation Agency (ARIS), Research core funding Grants No.~P1-0044, No.~N1-0273, and No.~J1-50005.
We gratefully acknowledge the Polish high-performance computing infrastructure PLGrid (HPC Centers: ACK Cyfronet AGH) for providing computer facilities and support within computational grant no. PLG/2025/018400.

%\bibliography{refs} 

\appendix
\section{Traces for $\gamma_0$ and $\gamma_1$. }
\label{ap:anal}
Let us start by decomposing the HSC Hamiltonian with onsite fields as a sum of $2-$site and $1-$site local operators. Leading to the expression for the Hamiltonian as $\hat{H}_{\mathrm{HSC}} \equiv \sum_{i=1}^{L-1}[ \hat{J}_{i,i+1}+\hat{V}_{i,i+1}]+\sum_{i=1}^L\hat{h}_i $, where $\hat{J}_{i,i+1}\equiv \frac{1}{2}(\hat{s}_i^+\hat{s}_{i+1}^-+\hat{s}_i^-\hat{s}_{i+1}^+)$, $\hat{V}_{i,i+1}=\hat{s}_i^z\hat{s}_{i+1}^z$ and $\hat{h}_i\equiv h_i\hat{s}_i^z$. Considering the zero magnetization Fock basis in the $Z-$basis, which has a dimension $D=L!/((L/2)!(L/2)!)$, we will estimate the various traces of interest associated with the derivations of $\gamma_0$ and $\gamma_1$. That is, $\mathrm{Tr}[\hat{s}_l^z]$, $\mathrm{Tr}[\hat{H}_{\mathrm{HSC}}]$, $\mathrm{Tr}[\hat{H}_{\mathrm{HSC}}^2]$ and $\mathrm{Tr}[\hat{s}_l^z \hat{H}_{\mathrm{HSC}}]$.

Working with the $Z-$base, it is immediately apparent that $\mathrm{Tr}[\hat{s}_l^z]$ is simply zero, so we start by calculating the trace of the Hamiltonian, $\mathrm{Tr}[\hat{H}_{\mathrm{HSC}}]$. The tunneling terms, $\hat{J}_{i,i+1}$, here will only have off-diagonal elements and thus do not contribute to the Hamiltonian trace. From $\mathrm{Tr}[\hat{s}_l^z]=0$ it is evident that the contributions from the onsite fields $\hat{h}_i$ are also zero. And the only terms that contribute are the interaction terms $\hat{V}_{i,i+1}$ which from the combinatorial restrictions associated with choosing two spins oriented in up/down directions within this magnetization restricted basis make a contribution to the basis as $\mathrm{Tr}[\hat{J}_{i,i+1}]=-{D}/(4(L-1))$. Thus, considering all terms for the HSC Hamiltonian with open boundary conditions, one gets: $\mathrm{Tr}[\hat{H}_{\mathrm{HSC}}]=-D/4$.

Next, considering the higher momentum of the Hamiltonian, $\mathrm{Tr}[\hat{H}_{\mathrm{HSC}}^2]$, one needs to consider pure squares of the local operators of the Hamiltonian and products of different local operators. The terms of the tunneling part that lead to products with other local operators are once again purely off-diagonal and do not contribute, so one is left with terms of the form $\hat{J}_{i,i+1}^2$ or $\hat{J}_{i,i+1}\hat{J}_{j,j+1}$ (with $i\neq j$) , here again all terms of the second form will only have off-diagonal elements and will have zero contribution. On expanding the trace associated with $\hat{J}_{i,i+1}^2$ one finds $\hat{J}_{i,i+1}^2=\mathrm{Tr}[(\hat{s}_i^x)^2(\hat{s}_{i+1}^x)^2+(\hat{s}_i^y)^2(\hat{s}_{i+1}^y)^2 + (\hat{s}_i^x)(\hat{s}_{i+1}^x) (\hat{s}_i^y)(\hat{s}_{i+1}^y)+(\hat{s}_i^y)(\hat{s}_{i+1}^y) (\hat{s}_i^x)(\hat{s}_{i+1}^x)]$.
The first two terms here are just scalars and will just have the contribution $\mathrm{Tr}[(\hat{s}_i^x)^2(\hat{s}_{i+1}^x)^2+(\hat{s}_i^y)^2(\hat{s}_{i+1}^y)^2 ]=D/8$, while for the second term with the Pauli commutation relations one can recover a local operator of the form $\hat{V}_{i,i+1}$; then making use of the combinatorial restrictions one finds $\mathrm{Tr}[(\hat{s}_i^y)(\hat{s}_{i+1}^y) (\hat{s}_i^x)(\hat{s}_{i+1}^x)]=D/(8(L-1))$. On summing over all terms one finds the contribution from the tunneling part as $\mathrm{Tr}[\sum_{i,j}\hat{J}_{i,i+1}\hat{J}_{j,j+1}]=DL/8$. While considering the interaction part the cross-term product with the onsite field operators will have no contribution as it will just produce $3-$site or $1-$site $\hat{s}_l^z$ operators, both of which have vanishing trace. Now, considering the pure products of the interaction part, one has terms of the form $\mathrm{Tr}[\hat{V}_{i,i+1}\hat{V}_{j,j+1}]$. For $i=j$ the trace will give the scalar contribution $D/16$. While, for $\vert i - j\vert =1$ the trace will take a form $\frac{1}{4}\mathrm{Tr}[\hat{s}_m^z\hat{s}_n^z]$ that makes a contribution $-D/(16(L-1))$. And for $\vert i-j \vert >1$ one has the full $4-$site operator of the form $\mathrm{Tr}[\hat{s}_i^z\hat{s}_{i+1}^z\hat{s}_j^z\hat{s}_{j+1}^z]$ that again, making use of combinatorics associated with evaluating the probability of choosing different choices of spins within this constrained basis, gives $\mathrm{Tr}[\hat{s}_i^z\hat{s}_{i+1}^z\hat{s}_j^z\hat{s}_{j+1}^z]=\frac{3{D}}{16(L-1)(L-3)}$. Summing over all indices the total contribution arising from the interacting part is $\mathrm{Tr}[\sum_{i,j}\hat{V}_{i,i+1}\hat{V}_{j,j+1}]=L/8-\frac{1}{16(L-1)}[2(L-2)]+\frac{3{D}}{16(L-1)(L-3)}[(L-2)(L-3)]$. And finally considering the field terms we find traces of the form $\mathrm{Tr}[h_i\hat{s}_i^zh_j\hat{s}_j^z]$ which when $i=j$ is a trace over a scalar that gives $Dh_i^2/4$ while for $i\neq j$ gives terms similar to $\hat{V}$ and giving the trace value, $\frac{Dh_ih_j}{4(L-1)}$. On summing over indices, $\mathrm{Tr}[\sum_{i,j}\hat{h}_i\hat{h}_j]=+1/4\sum_{i=1}^L h_i^2-1/(4(L-1))\sum_{i\neq j}h_ih_j$. Collating all contributions from the three terms of the Hamiltonian one finds: $\mathrm{Tr}[\hat{H}_{\mathrm{HSC}}^2]/D=(3L^2-3L-1)/(16(L-1))+1/4\sum_{i=1}^L h_i^2-1/(4(L-1))\sum_{i\neq j}h_ih_j$.

Finally, we look at $\mathrm{Tr}[\hat{H}\hat{s}_j^z]$. This trace just has a trace over local operators, which is up to a constant already dealt with while evaluating $\mathrm{Tr}[\hat{H}_{\mathrm{HSC}}^2]$.  Here, there are no contributions from the tunneling and the interaction part based on insights mentioned above. While from the field operators, one gets a trace over operators of the form $\mathrm{Tr}[h_i\hat{s}_i^z\hat{s}_j^z]$, which have already been evaluated up to a scalar above; one finds in the case $i=j$ a contribution of $Dh_i/4$, while for $i\neq j$ the contribution is $-({D}h_i)/(4(L-1))$. Summing over the indices, we find: $\mathrm{Tr}[\hat{H}_{\mathrm{HSC}}\hat{s}_j^z]/{D}=1/4 h_j-1/(4(L-1))\sum_{i\neq j}h_i$.

Replacing these traces in \eqref{eq:O6} and \eqref{eq:O7} gives the explicit linear form of $ \hat{s}_{l,{\|}}^z$ as
\begin{eqnarray}
 \hat{s}_{l,{\|}}^z&=&\frac{\left(\frac{1}{4}h_l-\frac{1}{4(L-1)}\sum_{i\neq l}h_i\right)\left[ \hat{H}_{\mathrm{HSC}}+1/4 \right]}{\frac{(3L^2-3L-1)}{(16(L-1))}+\frac{1}{4}\sum_ih_i^2+\frac{1}{4(L-1)}\sum_{i\neq j}h_ih_j-\left(\frac{1}{4}\right)^2}.\nonumber \\
 &~&
 \label{eq::s_zeqntot}
 \end{eqnarray}

\section{Density of states}
\label{ap:dens}

\begin{figure}
 \begin{center}
\includegraphics[width=1\linewidth]{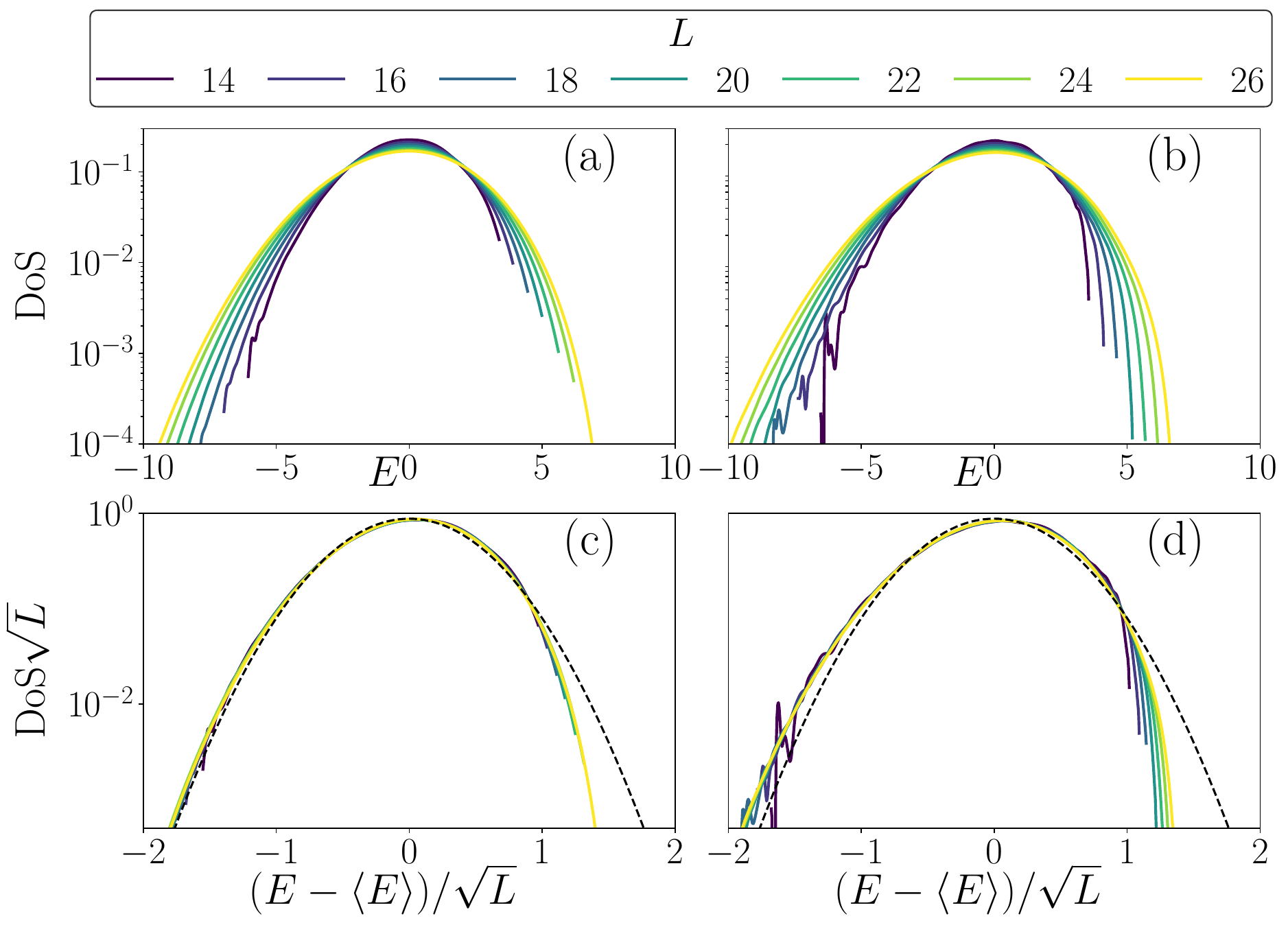}
\caption[]
{The density of states varying with $E$ for the HSC Hamiltonian with (a) random and (b) quasiperiodic disorder with different system sizes. The panels (c) and (d) correspond to the random and quasiperiodic disorder with the density of states and energies rescaled with a presumed standard deviation $\sqrt{L}$. The dashed line is the expected Gaussian form at very large system sizes as estimated from Appendix~\ref{ap:anal}. All plots are made for $W=0.5$ considering at least $10$ disorder realizations. The DoS are calculated based on a stochastic Kernel Polynomial approximation method using Chebyshev Polynomials as in \cite{Silver94,Silver96} expanded stochastically to the order of $200$ with $200$ random vectors.
}\label{fig:DOS_scale} 
 \end{center}
\end{figure}

The assumption often made in the context of ETH (see, e.g., \cite{Mierzejewski20}) is that the density of states (DoS) is a near-Gaussian with variance that depends linearly on the system size. The DoS is expected to take the particular form 
\begin{equation}
\rho(E)\approx \frac{\exp\left(-\frac{(E-\braket{E} )^2)}{2\sigma^2}\right)}{\sqrt{2\pi \sigma^2}}
    \label{eq::DOS}
\end{equation}
with $\sigma^2\propto L$.

The DoS for the HSC Hamiltonian with random and QP disorder at the disorder strength we consider, $W=0.5$, is approximately Gaussian with a slight skewness to low energies as shown in Fig.~ \ref{fig:DOS_scale}(a,b), where the DoS is calculated based on a stochastic Kernel Polynomial approximation method as in~\cite{Silver94,Silver96,Sierant20p}. Presuming a variance that linearly depends on the system size and making appropriate scalings one finds a leading single curve, as shown in Figs. \ref{fig:DOS_scale}(c,d), with additional system size corrections that approaches a Gaussian distribution with increasing system size.

An unskewed Gaussian distribution, as suggested by numericals, is expected for larger system sizes. This is because skeweness, $\gamma = \frac{k_3}{\sigma^3}$, where, $k_3$ is the third cumulant $k_3=\braket{(H-\braket{H})^3}$ and expected to grow at maximum as linearly with system size along with variance for a 1 dimensional ergodic Hamiltonian with short range interaction, goes as $\gamma \sim 1/\sqrt{L}$.

\section{Numerical details}
\label{ap: time}
The eigenstates and eigenvalues evaluated in the bulk of the spectrum were calculated for system sizes $L\leq18$ either with the standard exact diagonalization (\textbf{ED}) method for dense matrices or using the implementation of the shift-invert method in scipy~\cite{Scipy}. While for $L=20$ we used the \textit{polynomially filtered
exact diagonalization} (\textbf{POLFED}) method~\cite{Sierant20p}.

For the time dynamics considered in Fig.~\ref{fig:6}, we used the Chebyshev propagation scheme~\cite{Fehske08,Sierant21c}, which approaches the time-evolution operator, $U(\Delta t)=\exp(-iH\Delta t)$, by expansion in terms of the Chebyshev polynomials as
\begin{equation}
    U(\Delta t) \approx \left(J_0(a\Delta t)+2\sum_{k=1}^N(-i)^kJ_k(a\Delta t)T_k(\mathcal{H}) \right )
\end{equation}
where, $a=(E_{\max}-E_{\min})/2$, $b=(E_{\max}+E_{\min})/2$ and the Hamiltonian, $H$, is rescaled as $\mathcal{H}=\frac{1}{a}(H-b)$. Above, $J_{k}(t)$ is the Bessel function of order $k$ and $T_k(\mathcal{H})$ is the Chebyshev polynomial of order $k$. The order of expansion $N$ restricts the errors in the time evolution and is determined by a binary search to give values that makes the norm of $U(\Delta t)$ acting on a random normalized state deviating from unity not more than $10^{-13}$. For much longer times than we consider in this paper, rigorous numerical comparisons were made with ED for similar systems in~\cite{Sierant22c}.

\section{Disorder dependence}
\label{ap:new}

There is a simple argument to explain why the resemblance between the magnetization and disorder distributions shown in Fig.~\ref{fig:2} and Fig.~\ref{fig:3} becomes less and less pronounced as the disorder strength is increased.
Consider Eq.~\eqref{eq:szj}, we obtain the approximation
\begin{equation}
    s_j^z(E)\propto \frac{E-\braket{E}}{3L/4+\sum_ih_i^2}h_j\approx \frac{E-\braket{E}}{L(3/4+W^2\sigma^2)}h_j
    \label{eq::sig_approx}
\end{equation}
where $\sigma^2$ is the variance for unit disorder strength, which has a maximum value of $\sigma^2=1$ for the binary disorder. Using \eqref{eq::sig_approx},  one can show that the distribution for $s_j^z(E)$ for a fixed energy and system size gets broader (narrower) if $W<\sqrt{3}/(2\sigma)$ $\left(W>\sqrt{3}/(2\sigma)\right)$. This follows from the fact that taking a typical value of $h_j\approx W\sigma$ we obtain a dispersive curve in \eqref{eq::sig_approx}) which reaches extrema at $W=\sqrt{3}/(2\sigma)$.  For given values of $W$ in Figs.~2,3, assuming a fixed energy, \eqref{eq::sig_approx} one gets broadening distributions.

\begin{figure}
 \begin{center}
\includegraphics[width=\linewidth]{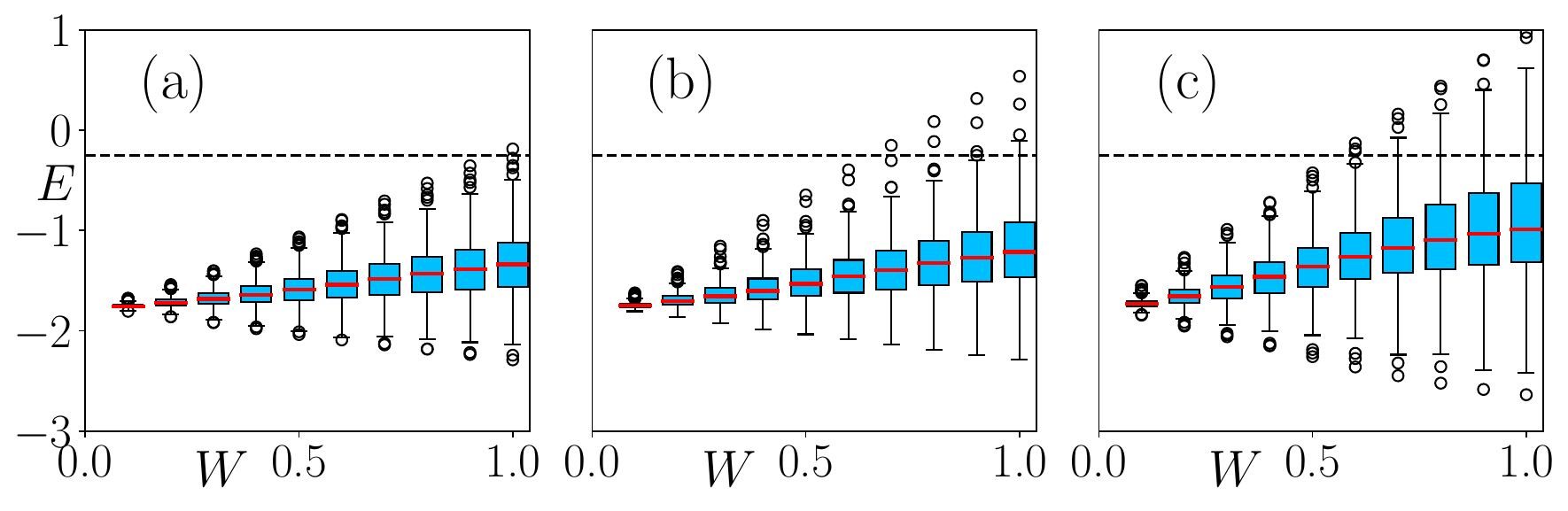}
\includegraphics[width=\linewidth]{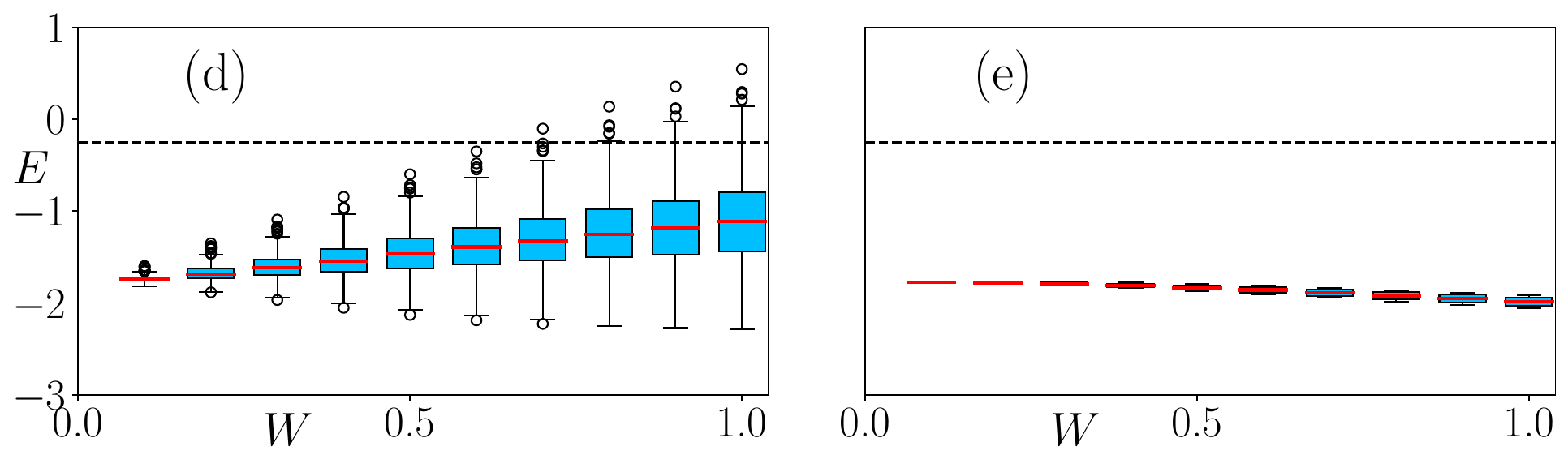}
\caption[]
{The distribution of energy values corresponding to $\epsilon =0.5$ over disorder realizations shown with standard box plots for a random (a), cosine (b), binary (c), tertiary (d), and triangle-wave quasiperiodic distribution. The dashed black line corresponds to $\braket{E}=-1/4$. All figures consider a system size $L=18$ with $500$ disorder realizations.
}\label{fig:ener_at_mid} 
 \end{center}
\end{figure}

The eigenstates for Fig.~\ref{fig:2} and Fig.~\ref{fig:3}  were chosen at the rescaled energy $\epsilon=0.5$, looking at the corresponding energy value $E$ with varying disorder realizations, we find in Fig.~\ref{fig:ener_at_mid} that for 
random, cosine, binary and tertiary disorders with increasing disorder strength, the energies approach the average energy, $\braket{E}=-1/4$,  with increasing fluctuations. This leads to the distribution of $s_j^z$ with $W$ becoming progressively smeared from the disorder distribution and eventually losing the peculiar disorder characteristics. For the quasiperiodic triangle wave distribution, there is comparatively less change in energy and fluctuations, leading to a more apparent broadening of the site-resolved magnetization
distribution.

\normalem

\end{document}